\documentclass[reprint,nofootinbib,showpacs,
preprintnumbers,amsmath,amssymb,aps]{revtex4-2}

\usepackage{braket}
\usepackage{parskip}
\usepackage{xcolor}
\usepackage{nicefrac}
\usepackage{pict2e}
\usepackage{xcolor}
\usepackage{mathtools}
\usepackage{yhmath}
\usepackage{diagbox}
\usepackage{arydshln}
\usepackage{bm}
\usepackage{graphbox}
\usepackage{amssymb}
\usepackage{mathtools}
\usepackage{enumitem}
\usepackage{scalerel,stackengine}
\usepackage{amsmath}
\usepackage{setspace}
\usepackage{tikz}
\usepackage{needspace}

\usepackage[colorlinks=true, linkcolor=magenta, citecolor=magenta, urlcolor=magenta]{hyperref}

\makeatletter
\renewcommand*\l@section[2]{%
  \ifnum \c@tocdepth > \z@
    \addpenalty \@secpenalty
    \addvspace {1.0em \@plus \p@ }%
    \setlength \@tempdima {2em}%
    \begingroup
      \parindent \z@
      \rightskip \@pnumwidth
      \parfillskip -\@pnumwidth
      \leavevmode
      \advance \leftskip \@tempdima
      \hskip -\leftskip
      \hypersetup{linkcolor=black}  
      \bfseries#1\nobreak{\normalfont\dotfill} 
      \nobreak
      \hb@xt@ \@pnumwidth {\hss \hyperlink{page.#2}{#2}\kern -\p@ \kern \p@ }
      \par
      \hypersetup{linkcolor=magenta} 
    \endgroup
  \fi
}
\makeatother

\makeatletter
\renewcommand*\l@subsection[2]{%
  \ifnum \c@tocdepth > 1
    \addpenalty \@secpenalty
    \begingroup
      \setlength \@tempdima {2.05em} 
      \parindent \z@
      \rightskip \@pnumwidth
      \parfillskip -\@pnumwidth
      \leavevmode
      \hskip \@tempdima 
      \llap{\Alph{subsection}\hskip 0.5em} 
      #1\nobreak
      \textcolor{magenta}{\normalfont\dotfill} 
      \hfil
      \nobreak
      \hb@xt@ \@pnumwidth {\hss \hyperlink{page.#2}{#2}\kern -\p@ \kern \p@ }
      \par
    \endgroup
  \fi
}
\makeatother

\makeatletter
\def\l@subsubsection#1#2{}  
\makeatother

\makeatletter
\def\l@f@section{}  
\makeatother

\def\toc@@font{}

\makeatletter
\def\toc@@font{\setlength{\parskip}{0.25em}}  
\makeatother

\makeatletter
\def\p@subsection{\thesection.}
\makeatother

\setlist[itemize]{leftmargin=18pt}
\setlist[enumerate]{leftmargin=18pt}

\newcommand\pig[1]{\scalerel*[5.5pt]{\Big#1}{%
  \ensurestackMath{\addstackgap[1.5pt]{\big#1}}}}
\newcommand\pigl[1]{\mathopen{\pig{#1}}}
\newcommand\pigr[1]{\mathclose{\pig{#1}}}

\newcommand*\circled[1]{\tikz[baseline=(char.base)]{%
            \node[shape=circle,fill=magenta!15,draw,inner sep=1pt] (char) {#1};}}

\newcommand\crule[3][black]{\textcolor{#1}{\rule{#2}{#3}}}

\newcommand\one{\unitlength=1pt
 \hspace{2.5pt}\begin{picture}(3,1)
 \linethickness{0.14mm}
 \put(-0.7,0.2){\line(0,1){6.5}}
 \put(0.7,0.2){\line(0,1){7.2}}
 \put(-2.1,0.2){\line(1,0){4.2}}
 \put(0.9,7.4){\line(-1,-0.4){3.2}}
 \end{picture}}
 
\newcommand\tub[4][]{\big\langle \unitlength=1pt
 \begin{picture}(20,1)
 \linethickness{0.18mm}
 \put(12,-4.2){\line(0,1){11}}
 \put(-0.5,2.7){\line(1,0){24}}
 \put(12,2.7){\color{gray}\circle*{3}}
 \put(10,8){$\scriptstyle #3$}
 \put(0.3,4.7){$\scriptstyle #2$}
 \put(18.7,4.7){$\scriptstyle #1$}
 \put(13.5,-3){\color{gray} $\scriptstyle #4$}
 \end{picture} \hspace{3.5pt} \big\rangle}
 
\newcommand\tuub[4][]{\big\langle \unitlength=1pt
 \begin{picture}(20,1)
 \linethickness{0.18mm}
 \put(20,-4.2){\line(0,1){11}}
 \put(-0.5,2.7){\line(1,0){40}}
 \put(20,2.7){\color{gray}\circle*{3}}
 \put(20,8){$\scriptstyle #3$}
 \put(0.3,4.7){$\scriptstyle #2$}
 \put(34,4.7){$\scriptstyle #1$}
 \put(21,-3.4){\color{gray} $\scriptstyle #4$}
 \end{picture} \hspace{18.5pt} \big\rangle}

\newcommand\htub[4][]{\big\langle \unitlength=1pt
 \begin{picture}(20,12)
 \linethickness{0.18mm}
 \put(9,-4.2){\line(0,1){11}}
 \put(15,4.4){\line(0,1){3.6}}
 \put(-0.5,2.7){\line(1,0){8}}
 \put(12,2.7){\line(1,0){12}}
 \put(12,2.7){\color{gray}\circle*{3}}
 \put(9,9.5){$\scriptstyle #3$}
 \put(0.3,4.7){$\scriptstyle #2$}
 \put(18.7,4.7){$\scriptstyle #1$}
 \put(13,-4){\color{gray} $\scriptstyle #4$}
 \put(9,-4.2){\line(30,6){4}}
 \put(9,6.8){\line(30,6){6}}
 \end{picture} \hspace{3.5pt} \big\rangle}
 
\newcommand\htuub[4][]{\big\langle \unitlength=1pt
 \begin{picture}(20,12)
 \linethickness{0.18mm}
 \put(9,-4.2){\line(0,1){11}}
 \put(15,4.4){\line(0,1){3.6}}
 \put(-0.5,2.7){\line(1,0){8}}
 \put(12,2.7){\line(1,0){20}}
 \put(12,2.7){\color{gray}\circle*{3}}
 \put(9,10){$\scriptstyle #3$}
 \put(0.3,4.7){$\scriptstyle #2$}
 \put(26.5,4.7){$\scriptstyle #1$}
 \put(13,-4){\color{gray} $\scriptstyle #4$}
 \put(9,-4.2){\line(30,6){4}}
 \put(9,6.8){\line(30,6){6}}
 \end{picture} \hspace{11.5pt} \big\rangle}

\newcommand\chtub[3][]{\big\langle \unitlength=1pt
 \begin{picture}(20,12)
 \linethickness{0.18mm}
 \put(9,-4.2){\line(0,1){11}}
 \put(15,-3.02){\line(0,1){11}}
 \put(-0.5,2.7){\line(1,0){8}}
 \put(16.8,2.7){\line(1,0){7.2}}
 \put(9.7,9.5){$\scriptstyle #3$}
 \put(0.3,4.7){$\scriptstyle #2$}
 \put(18.7,4.7){$\scriptstyle #1$}
 \put(9,-4.2){\line(30,6){6}}
 \put(9,6.8){\line(30,6){6}}
 \end{picture} \hspace{3.5pt} \big\rangle}
 
\newcommand\uhtuub[4][]{\big\langle \unitlength=1pt
 \begin{picture}(20,12)
 \linethickness{0.18mm}
 \put(9,-4.2){\line(0,1){11}}
 \put(15,4.4){\line(0,1){3.6}}
 \put(-0.5,2.7){\line(1,0){8}}
 \put(12,2.7){\line(1,0){15}}
 \put(12,2.7){\color{gray}\circle*{3}}
 \put(9,10){$\scriptstyle #3$}
 \put(0.3,4.7){$\scriptstyle #2$}
 \put(21.5,4.7){$\scriptstyle #1$}
 \put(13,-4){\color{gray} $\scriptstyle #4$}
 \put(9,-4.2){\line(30,6){4}}
 \put(9,6.8){\line(30,6){6}}
 \end{picture} \hspace{6.7pt} \big\rangle}
 
\newcommand\htubb[4][]{\big\langle \unitlength=1pt
 \begin{picture}(20,1)
 \linethickness{0.18mm}
 \put(9,-4.2){\line(0,1){11}}
 \put(15,-3){\line(0,1){4}}
 \put(15,4.4){\line(0,1){3.6}}
 \put(-0.5,2.7){\line(1,0){8}}
 \put(12,2.7){\line(1,0){12}}
 \put(12,2.7){\color{gray}\circle*{3}}
 \put(10,10.5){$\scriptstyle #3$}
 \put(0.3,4.7){$\scriptstyle #2$}
 \put(18.7,4.7){$\scriptstyle #1$}
 \put(13.5,-3){\color{gray} $\scriptstyle #4$}
 \put(9,-4.2){\line(30,6){6}}
 \put(9,6.8){\line(30,6){6}}
 \end{picture} \hspace{3.5pt} \big\rangle}

\begin{document}

\setcounter{tocdepth}{2}

\title{Unitary Categorical Symmetries}

\author{T. Bartsch}
 
\affiliation{Department of Mathematical Sciences, \\ Durham University, United Kingdom}

\begin{abstract}
Global invertible symmetries act unitarily on local observables or states of a quantum system. In this note, we aim to generalise this statement to non-invertible symmetries by considering unitary actions of higher fusion category symmetries $\mathcal{C}$ on twisted sector local operators. We propose that the latter transform in $\ast$-representations of the tube algebra associated to $\mathcal{C}$, which we introduce and classify using the notion of higher $S$-matrices of higher braided fusion categories.
\end{abstract}

\maketitle

\tableofcontents

\section{Introduction}

According to Wigner's theorem \cite{Wigner1931}, invertible global symmetries act unitarily (or anti-unitarily) on local observables or states of a quantum system. It is natural to ask how this generalises to the action of non-invertible symmetries described by (higher) fusion categories $\mathcal{C}$ \cite{Etingof2002,Gaiotto:2014kfa,Bhardwaj:2022yxj,Shao:2023gho,Schafer-Nameki:2023jdn}. In this note, we try to answer this question in two steps:
\begin{enumerate}
\item We consider the \textit{tube algebra} $\text{Tube}(\mathcal{C})$ \cite{Ocneanu2016ChiralityFO,Bullivant2019,Bartsch2023a} associated to $\mathcal{C}$ and show that it canonically possesses the structure of a C*-algebra, provided that the symmetry category $\mathcal{C}$ is unitary in an appropriate sense.
\item We propose that twisted sector local operators transform in $\ast$-representations of $\text{Tube}(\mathcal{C})$, which we construct and classify using the \textit{Symmetry TFT} \cite{Gaiotto2021,Apruzzi2023,Freed2022} and its associated \textit{higher $S$-matrices} \cite{JohnsonFreyd2021,Reutter2022,Johnson-Freyd:2021chu}. 
\end{enumerate}
While the above is well known in $D=2$ \cite{Neshveyev2018,Izumi2000,Lin2023}, we provide a general construction for arbitrary $D \geq 2$ including several examples in $D=2,3$.


\subsection{Background}

In Euclidean (Wick-rotated) quantum field theory, unitarity manifests itself in the principle of \textit{reflection positivity} \cite{Osterwalder1973,Osterwalder1975,Jaffe:2018ftu,Freed:2016rqq}. Concretely, upon fixing an affine hyperplane $\Pi$ in $D$-dimensional spacetime ($D \geq 2$), \textit{reflection} states that reflecting the operator content of a correlation function about $\Pi$ is equivalent to complex conjugating the correlation function,
\begin{equation}
\vspace{-5pt}
\begin{gathered}
\includegraphics[height=1.57cm]{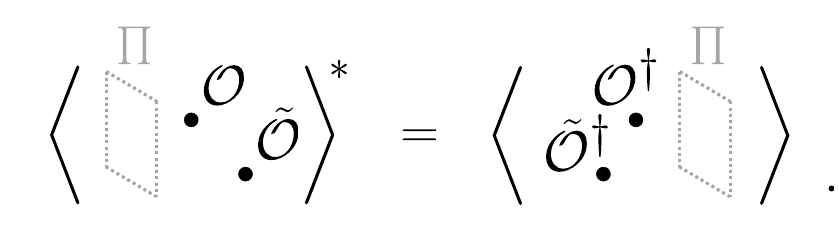}
\end{gathered}
\end{equation}
Here, $\mathcal{O}^{\dagger}$ denotes the operator that is obtained by reflecting an operator $\mathcal{O}$ about $\Pi$. \textit{Positivity} states that
\begin{equation}
\vspace{-5pt}
\begin{gathered}
\includegraphics[height=1.57cm]{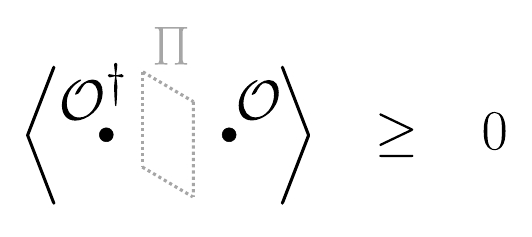}
\end{gathered}
\end{equation}
for all operators $\mathcal{O}$, with equality if and only if $\mathcal{O}=0$.

Now suppose that a given quantum field theory $\mathfrak{T}$ admits finite generalised symmetries described a fusion $(D-1)$-category $\mathcal{C}$, which we call the \textit{symmetry category} of $\mathfrak{T}$ in what follows. The latter captures topological defects in $\mathfrak{T}$ of all possible dimensions, where a $p$-morphism in $\mathcal{C}$ corresponds to a codimension-$(p+1)$ topological symmetry defect in the theory ($p=0,...,D-1$):
\begin{equation}
\vspace{-5pt}
\begin{gathered}
\includegraphics[height=1.61cm]{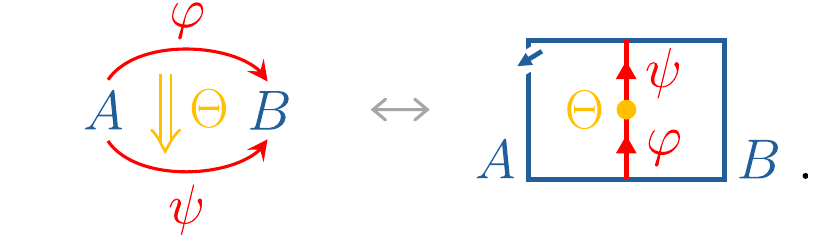}
\end{gathered}
\end{equation}
Given a twisted sector local operator $\mathcal{O}$ attached to a topological line defect\footnote{Here, we denote by $\Omega(\mathcal{C}) := \text{End}_{\mathcal{C}}(\mathbf{1})$ the endomorphisms of the monoidal unit $\mathbf{1} \in \mathcal{C}$. Inductively, $\Omega^n(\mathcal{C}) := \Omega(\Omega^{n-1}(\mathcal{C}))$.} $\mu \in \Omega^{D-2}(\mathcal{C})$, symmetry defects $A \in \mathcal{C}$ can act on $\mathcal{O}$ via linking,
\begin{equation}
\vspace{-5pt}
\begin{gathered}
\includegraphics[height=1.57cm]{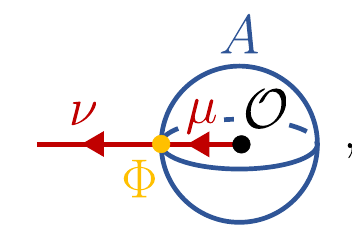}
\end{gathered}
\end{equation}
mapping $\mathcal{O}$ to a local operator in a potentially different twisted sector $\nu \in \Omega^{D-2}(\mathcal{C})$. Configurations of the above type then generate the \textit{tube algebra}\footnote{In the language of \cite{Bartsch2023a}, the above is the $S^{D-1}\hspace{1pt}$--$\hspace{1pt}$tube algebra associated to the linking of local operators with codimension-one symmetry defects placed on a $(D-1)$-sphere. For the purposes of this note, we will simply call it the \textit{tube algebra} in what follows.} $\text{Tube}(\mathcal{C})$ of $\mathcal{C}$, which is a finite-dimensional complex associative algebra whose elements we denote by
\begin{equation}
\htub[\mu]{\nu}{A}{\Phi} \; \in \; \text{Tube}(\mathcal{C})
\end{equation}
in what follows (see \cite{Lin2023} for a discussion of the tube algebra and its physical applications in two dimensions and \cite{Bartsch2023a} for generalisations to higher $D$). Using this, the unit of $\text{Tube}(\mathcal{C})$ can be written as\footnote{Here, we denote by $\pi_0(\mathcal{B})$ the set of connected components of a (higher) fusion category $\mathcal{C}$, i.e. the set of simple objects in $\mathcal{C}$ modulo the existence of a non-zero morphism between them. Furthermore, we set $\pi_n(\mathcal{C}) := \pi_0(\Omega^n(\mathcal{C}))$. We often call $\pi_1(\mathcal{C})$ the \textit{fundamental hypergroup} of $\mathcal{C}$.}
\begin{equation}
\one \quad =  \sum_{[\mu] \, \in \, \pi_{D-2}(\mathcal{C})} \! \uhtuub[\mu]{\mu}{\hspace{0.5pt}\mathbf{1}}{\hspace{1.5pt} \text{Id}_{\hspace{0.1pt}\mu}} \; ,
\end{equation}
where $\mathbf{1} \in \mathcal{C}$ denotes the monoidal unit in $\mathcal{C}$. Twisted sector local operators $\mathcal{O}$ then transform in linear representations of the tube algebra, which can be constructed using a operator-state map\footnote{For generic $\mathfrak{T}$, this map will be non-surjective. By abuse of notation, we will henceforth use $\mathcal{H}_{\mu}$ to denote the subsector of the $\mu$-tiwsted Hilbert space that is given by the image of the operator-state map. Furthermore, since we are interested in the action of finite symmetries on these spaces, we will without loss of generality assume the $\mathcal{H}_{\mu}$ to be finite-dimensional.} as follows: By computing the half-space correlation function
\begin{equation}
\vspace{-5pt}
\begin{gathered}
\includegraphics[height=1.56cm]{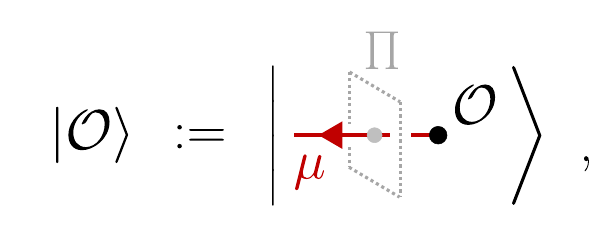}
\end{gathered}
\end{equation}
we obtain a state $| \mathcal{O} \rangle \in \mathcal{H}_{\mu}$ in the $\mu$-twisted Hilbert space of the theory, which is acted upon by elements of the tube algebra via
\begin{equation}
\vspace{-5pt}
\begin{gathered}
\includegraphics[height=1.56cm]{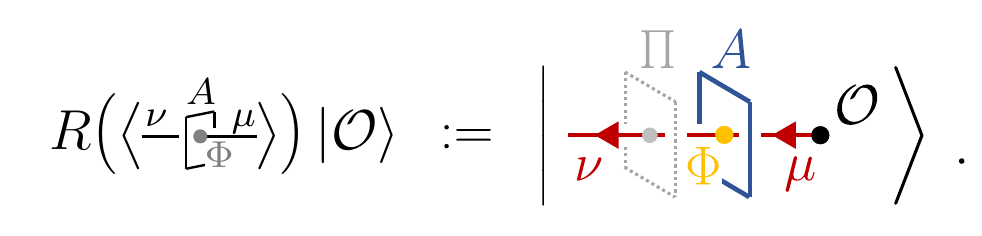}
\end{gathered}
\end{equation}
This induces linear maps $R\pigl(\htub[\mu]{\nu}{A}{\Phi}\pigr): \mathcal{H}_{\mu} \to \mathcal{H}_{\nu}$, which assemble into a linear representation
\begin{equation}
R: \; \text{Tube}(\mathcal{C}) \, \to \, \text{End}(\mathcal{H})
\end{equation}
of $\text{Tube}(\mathcal{C})$ on the total Hilbert space $\mathcal{H} = \bigoplus_{\mu} \mathcal{H}_{\mu}$. In order for this representation to be compatible with reflection positivity, we require that
\begin{equation}
\vspace{-5pt}
\begin{gathered}
\includegraphics[height=4.1cm]{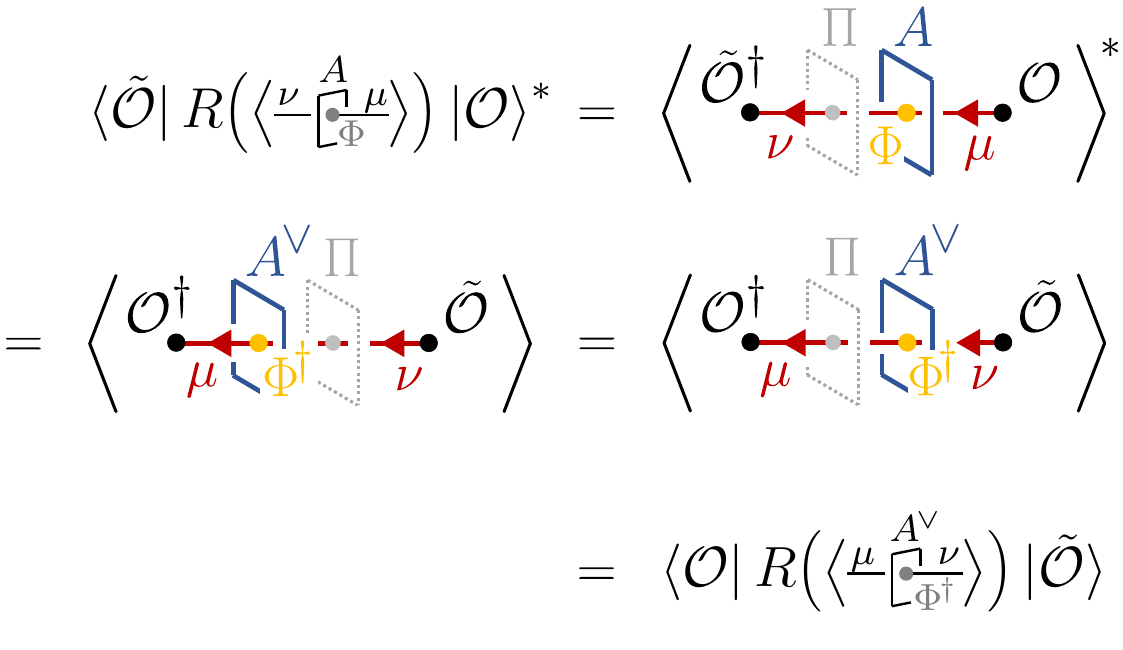}
\end{gathered}
\end{equation}
for all $\mathcal{O}$ and $\tilde{\mathcal{O}}$, where we used that reflection about $\Pi$ maps the codimension-one symmetry $A$ defect to its orientation reversal $A^{\vee}$. As a result, we see that
\begin{equation}
\label{eq-star-condition}
R\pigl(\htub[\mu]{\nu}{A}{\Phi}\pigr)^{\dagger} \,\; \stackrel{!}{=} \,\; R\pigl(\hspace{1pt}\htub[\mu]{\nu}{A}{\Phi}^{\ast}\pigr) \; ,
\end{equation}
where the prescription
\begin{equation}
\label{eq-tube-alg-invol}
\htub[\mu]{\nu}{A}{\Phi}^{\ast} \; := \; \htub[\nu]{\mu}{\hspace{-0.5pt}A^{\hspace{-0.6pt}\vee}}{\hspace{-0.7pt}\raisebox{-1.1pt}{\hspace{1.2pt}$\scriptstyle \Phi^{\hspace{-0.5pt}\dagger}$}}
\end{equation}
defines an antilinear involution\footnote{An \textit{involution} on a set $X$ is a map $\ast: X \to X$ such that $\ast^2 = \text{id}_X$.} $\ast: \text{Tube}(\mathcal{C}) \to \text{Tube}(\mathcal{C})$ on the tube algebra. Since for $\mu \in \Omega^{D-2}(\mathcal{C})$ simple,
\begin{equation}
\label{eq-positive-functional}
F\pigl(\htub[\mu]{\nu}{A}{\Phi}\pigr) \; := \; \delta_{A,\mathbf{1}} \cdot \delta_{\mu,\nu} \cdot \Phi
\end{equation}
induces a faithful positive functional\footnote{Here, we used the fact that for $\mu \in \Omega^{D-2}(\mathcal{C})$ simple, we have that $\Phi \in (D-1)\text{-End}_{\mathcal{C}}(\mu) \cong \mathbb{C}$.} $F: \text{Tube}(\mathcal{C}) \to \mathbb{C}$, this shows that $\text{Tube}(\mathcal{C})$ is a C*-algebra \cite{Izumi2000}. Condition (\ref{eq-star-condition}) then means $R$ is a \textit{$\ast$-representation} of $\text{Tube}(\mathcal{C})$, which forms a natural generalisation of the notion of a unitary representation of a finite group. The aim of this note is to classify all irreducible $\ast$-representations of the tube algebra for a given symmetry category $\mathcal{C}$.

\subsection{Overview}
\label{ssec-overview}

In a unitary quantum field theory with finite internal symmetry group $G$, local operators transform in unitary representations of $G$, which correspond to group homomorphisms
\begin{equation}
R: \; G \, \to \, U(\mathcal{H})
\end{equation}
from $G$ into the unitary group of a Hilbert space $\mathcal{H}$. Equivalently, by linear continuation, we may view $R$ as a $\ast$-representation of the group algebra $\mathbb{C}[G]$ of $G$,
\begin{equation}
\label{eq-group-rep}
R: \; \mathbb{C}[G] \, \to \, \text{End}(\mathcal{H}) \, ,
\end{equation}
where the involution $\ast$ on $\mathbb{C}[G]$ is the antilinear continuation of the inversion map $g \mapsto g^{-1}$.

As motivated in the previous subsection, upon replacing the finite group $G$ with a generalised (higher) fusion category symmetry $\mathcal{C}$, the natural generalisation of (\ref{eq-group-rep}) is a $\ast$-representation of the tube algebra of $\mathcal{C}$ on the Hilbert space $\mathcal{H}$ of (genuine and twisted sector) local operators, where the involution $\ast$ on $\text{Tube}(\mathcal{C})$ is as in (\ref{eq-tube-alg-invol}). In particular, this requires the symmetry category $\mathcal{C}$ to be equipped with the following:
\begin{enumerate}[label=\textbf{\arabic*}.]
\item \textbf{Duals:} For each symmetry defect $A \in \mathcal{C}$ there exists a \textit{dual defect} $A^{\vee} \in \mathcal{C}$ that corresponds to the orientation reversal of $A$ obtained by ``bending" the topological defect $A$ around, 
\begin{equation}
\vspace{-5pt}
\begin{gathered}
\includegraphics[height=1.2cm]{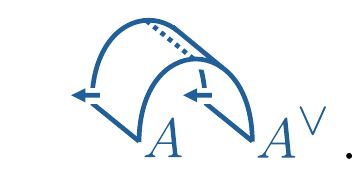}
\end{gathered}
\end{equation}
\item \textbf{Daggers:} For each local junction $\Theta$ in $\mathcal{C}$, there exists its \textit{dagger} $\Theta^{\dagger}$ which corresponds to the reflection of $\Theta$ about a fixed hyperplane\footnote{In general, for an $(\infty,n)$-category with all adjoints, different \linebreak $\dagger$-structures are parameterised by different choices of subgroup $\mathfrak{G} \subset \text{Aut}(\text{AdjCat}_{(\infty,n)}) \cong \text{PL}(n)$ \cite{Ferrer:2024vpn}. In our case, we choose $\mathfrak{G} = \mathbb{Z}_2$ implementing involutory reflections only at the top level of morphisms. In this way, we inductively view a $\dagger$-$n$-category as an $n$-category enriched in $\dagger$-$(n-1)$-categories.},
\begin{equation}
\vspace{-5pt}
\begin{gathered}
\includegraphics[height=1.5cm]{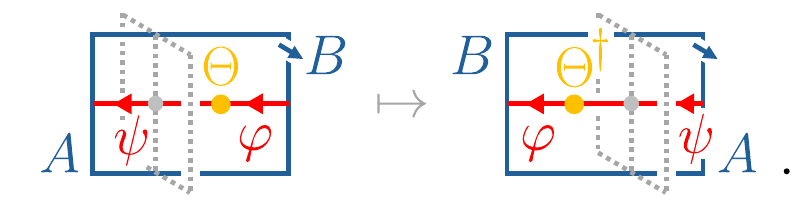}
\end{gathered}
\end{equation}
\end{enumerate}
Furthermore, we require the above structures to be compatible with one another in an appropriate sense in order for the involution $\ast$ on $\text{Tube}(\mathcal{C})$ to be well-defined.

A convenient tool to characterise and classify irreducible $\ast$-representations of the tube algebra is the so-called \textit{sandwich construction} \cite{Gaiotto2021,Apruzzi2023,Freed2022}. In this picture, a $D$-dimensional theory $\mathfrak{T}$ with generalised symmetry $\mathcal{C}$ is viewed as an interval compactification of a $(D+1)$-dimensional topological theory (called the \textit{Symmetry TFT}) with canonical topological boundary condition $\mathbb{B}_{\mathcal{C}}$ on the left and a physical (typically non-topological) boundary condition $\mathbb{B}_{\hspace{0.7pt}\mathfrak{T}}$ on the right. Twisted sector local operators $\mathcal{O}$ in $\mathfrak{T}$ then correspond to junctions between a left boundary line $\mu$ and a bulk line $\rho$ stretched between the two boundaries of the $(D+1)$-dimensional bulk:  
\begin{equation}
\vspace{-5pt}
\begin{gathered}
\includegraphics[height=3cm]{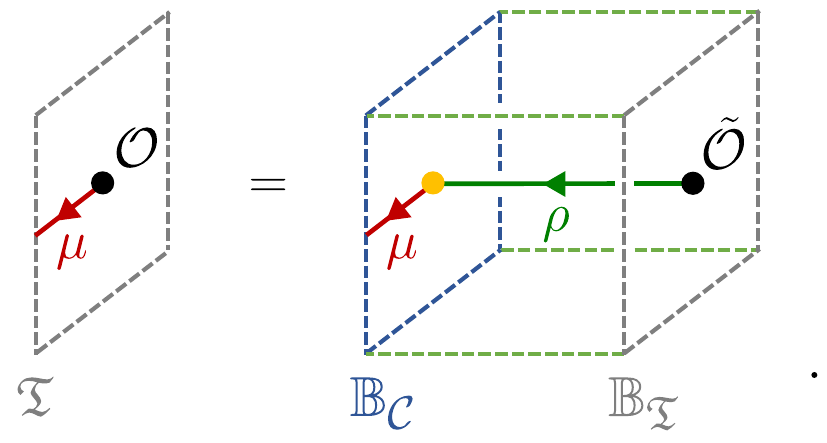}
\end{gathered}
\end{equation}
Now suppose that we link a twisted sector local operator $\mathcal{O}$ with a symmetry defect $U \in \mathcal{C}$ that can be pulled into the $(D+1)$-dimensional bulk: 
\begin{equation}
\label{eq-sandwich-construction-2}
\vspace{-5pt}
\begin{gathered}
\includegraphics[height=3.25cm]{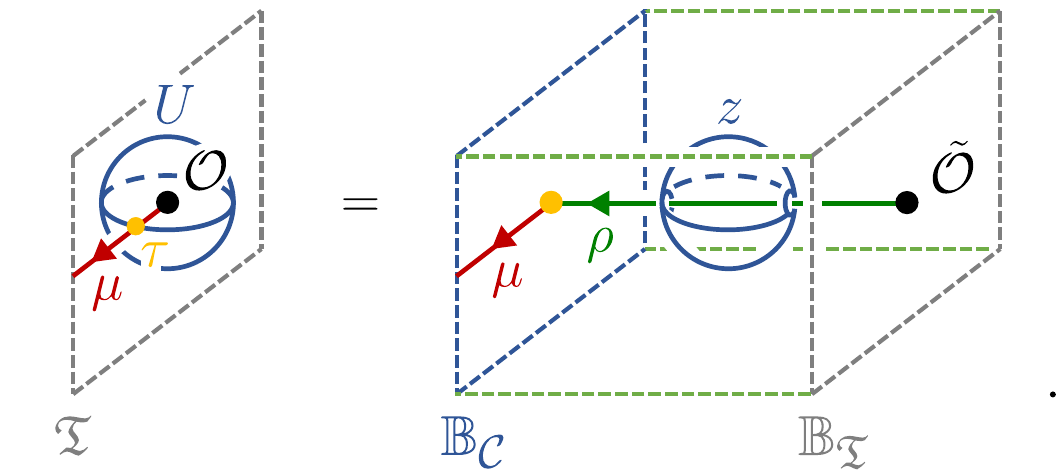}
\end{gathered}
\end{equation}
Since the latter hosts a Turaev-Viro TFT \cite{Turaev1992,Turaev1994,Barrett1996,Barrett1999} based on the symmetry category $\mathcal{C}$ (or higher-dimensional analogues thereof \cite{Douglas:2018qfz}), this necessitates $U$ to form part of a $D$-tupel $z = (U,\tau,...) \in \mathcal{Z}(\mathcal{C})$ in the Drinfeld center of $\mathcal{C}$. In particular, the datum $\tau$ corresponds to a half-braiding for $U$, which allows us to commute $U$ past any other topological defect in $\mathcal{C}$. For brevity, we denote the corresponding tube algebra element by
\begin{equation}
\htuub[\mu]{\mu}{\hspace{0.5pt}U}{\hspace{0.5pt}\raisebox{1pt}{$\scriptstyle \tau_{\hspace{0.5pt}U,\mu}$}}  \; =: \; \chtub[\mu]{\mu}{z}
\end{equation}
in what follows. This has the property that
\begin{equation}
\label{eq-central-tubes-commuting-prop}
\htub[\mu]{\nu}{A}{\Phi} \, \circ \, \chtub[\mu]{\mu}{z} \;\; = \;\; \chtub[\nu]{\nu}{z} \, \circ \, \htub[\mu]{\nu}{A}{\Phi}
\end{equation}
for all $\htub[\mu]{\nu}{A}{\Phi} \in \text{Tube}(\mathcal{C})$, with involution given by\footnote{Here, we implicitly assume that $z = (U,\tau,...)$ lies in the \textit{unitary Drinfeld centre} $\mathcal{Z}^{\dagger}(\mathcal{C})$, meaning that the top components associated to the half-braiding $\tau$ are unitary: $(\tau_{U,\mu})^{\dagger} = (\tau_{U,\mu})^{-1}$ for all $\mu \in \Omega^{D-2}(\mathcal{C})$. While $\mathcal{Z}^{\dagger}(\mathcal{C}) \cong \mathcal{Z}(\mathcal{C})$ in $D=2$ \cite{Galindo2014}, we expect this to hold also in $D>2$.}
\begin{equation}
\label{eq-central-tubes-conjugation}
\chtub[\mu]{\mu}{z}^{\ast} \; = \; \chtub[\mu]{\mu}{z^{\hspace{-1pt}\vee}} \; .
\end{equation}
From (\ref{eq-sandwich-construction-2}), it is then clear that the action of $\chtub[\mu]{\mu}{z}$ on $\mathcal{O}$ is completely determined by the linking of the topological defects $z$ and $\rho$ in the $(D+1)$-dimensional bulk, i.e.
\begin{equation}
\label{eq-diagonal-tube-action}
R\pigl( \chtub[\mu]{\mu}{z} \pigr) \, |\mathcal{O}\rangle \;\, = \;\, d_z \cdot S_{z,\rho} \cdot |\mathcal{O}\rangle \; ,
\end{equation}
where we abbreviated $d_z := \text{dim}(z)$ and defined the multiplicative factor\footnote{Note that this picture schematically represents the linking of an $S^{D-1}$ and an $S^1$ in a $(D+1)$-dimensional ambient space.}
\begin{equation}
\vspace{-5pt}
\begin{gathered}
\includegraphics[height=1.345cm]{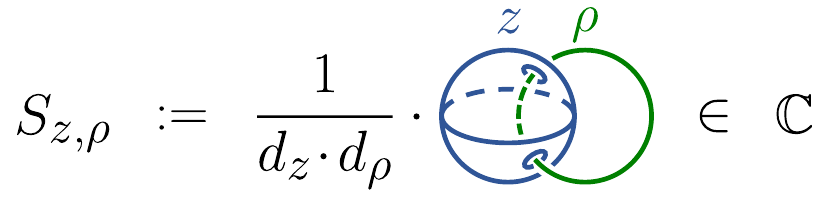}
\end{gathered}
\end{equation}
indexed by $z \in \mathcal{Z}(\mathcal{C})$ and $\rho \in \Omega^{D-2}(\mathcal{Z}(\mathcal{C}))$. The collection of these factors then descends to a well-defined pairing \cite{JohnsonFreyd2021,Reutter2022}
\begin{equation}
S:\; \; \pi_0(\mathcal{Z}(\mathcal{C})) \, \times \, \pi_{D-2}(\mathcal{Z}(\mathcal{C})) \;\, \to \;\, \mathbb{C} \; ,
\end{equation}
which we call the \textit{$S$-matrix} of $\mathcal{Z}(\mathcal{C})$ in what follows\footnote{For a general braided fusion $n$-category $\mathcal{B}$, there exists a whole collection of (generalised) $S$-matrices \cite{JohnsonFreyd2021,Reutter2022}
\begin{equation}
S^{(k)}: \; \pi_k(\mathcal{B}) \, \times \, \pi_{n-1-k}(\mathcal{B}) \; \to \; \mathbb{C} \; ,
\end{equation}
where $k=0,...,n-1$. For $n=1$, this reduces to the ordinary $S$-matrix for braided fusion categories \cite{Etingof2017}. In this note, we consider the case $n=D-1$ and $k=0$ with $\mathcal{B} = \mathcal{Z}(\mathcal{C})$ the Drinfeld centre of the symmetry $(D-1)$-category $\mathcal{C}$.}. This matrix has the following properties:
\begin{enumerate}[label=\protect\circled{\arabic*}]
\item It is invertible, so in particular we have that
\begin{equation}
|\pi_0(\mathcal{Z}(\mathcal{C}))| \; = \; |\pi_{D-2}(\mathcal{Z}(\mathcal{C}))| \; .
\end{equation}
\item It satisfies $S_{\hspace{0.5pt}z^{\vee}\!,\hspace{1.2pt}\rho} = (S_{z,\rho})^{\ast}$ for all $z$ and $\rho$.
\item It satisfies the Verlinde formula
\begin{equation}
S_{x,\rho} \hspace{1pt} \cdot \hspace{1pt} S_{y,\rho} \;\;  =  \, \sum_{[z] \hspace{1pt} \in \hspace{1pt} \pi_0} \! N_{xy}^z \hspace{1pt} \cdot \hspace{1pt} S_{z,\rho}
\end{equation}
for some unique coefficients $N_{xy}^z \in \mathbb{C}$.
\end{enumerate}
From a physical perspective, the (generically non-integer) coefficients $N_{xy}^z$ capture the tube algebra products
\begin{equation}
\label{eq-central-tubes-composition}
\chtub[\mu]{\mu}{x} \hspace{1pt} \circ \hspace{1pt} \chtub[\mu]{\mu}{y} \;\; = \sum_{[z] \hspace{1pt} \in \hspace{1pt} \pi_0} \!\! \frac{d_x \hspace{-1pt} \cdot \hspace{-1pt} d_y}{d_z} \cdot N_{xy}^z \cdot \chtub[\mu]{\mu}{z} \, .
\end{equation}
From a mathematical perspective, they define a hypergroup structure on the set $\pi_0(\mathcal{Z}(\mathcal{C}))$ of connected components of the Drinfeld centre of $\mathcal{C}$ \cite{Reutter2022}:

\textbf{Definition} \cite{Bloom1995,Sunder2003,Reutter2022,Kaidi2024}\textbf{:} A \textit{$\mathbb{C}$-hypergroup} is a finite set $X$ with an involution $\vee: X \to X$, a distinguished element $e \in X$ and a collection $\lbrace N_{xy}^z \in \mathbb{C}\rbrace_{x,y,z \hspace{2pt} \in \hspace{1pt} X}$ such that
\begin{enumerate}
\item setting $x \cdot y := \sum_z N^z_{xy} \cdot z$ defines an associative algebra structure on $\mathbb{C}[X]$,
\item the distinguished element $e \in X$ acts as a unit for the above algebra,
\item the continuation of $\vee$ to $\mathbb{C}[X]$ defines an algebra anti-isomorphism,
\item for all $x,y \in X$, we have $N_{xy}^e \neq 0 \; \Leftrightarrow \; y = x^{\vee}$,
\item for all $x,y \in X$, we have $\sum_z N_{xy}^z = 1$.
\end{enumerate}

For the purposes of this note, the utility of the of the $S$-matrix stems from the fact that it captures diagonal actions of the tube algebra on twisted sector local operators as in (\ref{eq-diagonal-tube-action}). In fact, the latter are sufficient to determine the entire action of the tube algebra (for $D=2$, this was spelled out in detail by the authors of \cite{Lin2023}, and the following discussion is a straight-up generalisation of their arguments to $D>2$): Given a simple left boundary line $\mu \in \Omega^{D-2}(\mathcal{C})$ and a simple line $\rho \in \Omega^{D-2}(\mathcal{Z}(\mathcal{C}))$ in the Symmetry TFT, we can define an associated tube algebra element
\begin{equation}
e^{\mu}_{\rho} \;\; :=  \, \sum_{[z] \hspace{1pt} \in \hspace{1pt} \pi_0} \frac{1}{d_z} \cdot (S^{-1})_{\rho,z} \cdot \chtub[\mu]{\mu}{z} \; ,
\end{equation}
where we used the invertibility $\circled{1}$ of the $S$-matrix. Using the Verlinde formula $\circled{3}$ and the composition rule (\ref{eq-central-tubes-composition}), one can then check that the above elements satisfy
\begin{equation}
e_{\rho}^{\mu} \, \circ \, e_{\sigma}^{\nu} \,\; = \,\; \delta_{\rho,\sigma} \hspace{1pt}\cdot\hspace{1pt} \delta^{\mu,\nu} \hspace{1pt}\cdot\hspace{1pt} e_{\rho}^{\mu} \;\, .
\end{equation}
Furthermore, using (\ref{eq-central-tubes-conjugation}) and $\circled{2}$ one finds $(e_{\rho}^{\mu})^{\ast} = e_{\rho}^{\mu}$. As a result, using (\ref{eq-central-tubes-commuting-prop}) we see that
\begin{equation}
\label{eq-min-cen-idems}
e_{\rho} \;\;\, := \sum_{[\mu] \hspace{1pt} \in \hspace{1pt} \pi_{D-2}} \! e_{\rho}^{\mu}
\end{equation}
defines a self-adjoint central idempotent (projector) in $\text{Tube}(\mathcal{C})$ for every simple $\rho \in \Omega^{D-2}(\mathcal{Z}(\mathcal{C}))$, which is \textit{minimal} in the sense that $e_{\rho} \neq e_1 + e_2$ for two non-zero central idempotents $e_1$ and $e_2$. Since the set of minimal projectors is in 1:1-correspondence with the irreducible $\ast$-representations of the tube algebra, this establishes an equivalence (see \cite{Popa2017,Gruen2021,HARDIMAN2020} for $D=2$)
\begin{equation}
\label{eq-classif-star-reps}
\text{Rep}^{\dagger}(\text{Tube}(\mathcal{C})) \;\, \cong \;\, \Omega^{D-2}(\mathcal{Z}^{\dagger}(\mathcal{C})) \; ,
\end{equation}
where the right hand side denotes the category of $\ast$-repre-sentations of $\text{Tube}(\mathcal{C})$ and intertwiners between them.

The aim of this note is to use the equivalence (\ref{eq-classif-star-reps}) to classify unitary actions of (higher) fusion category symmetries on twisted sector local operators in two and three spacetime dimensions. We provide an explicit construction of the tube algebra and its canonical $\ast$-structure as well as a variety of examples in sections \ref{sec-two-dimensions} ($D=2$) and \ref{sec-three-dimensions} ($D=3$), respectively. 

As a by-product, we find that the $S$-matrix associated to the Drinfeld centre of a finite 2-group symmetry \cite{Baez2004,Benini:2018reh} $\mathcal{G} = A[1] \rtimes G$ in three dimensions is given by
\begin{equation}
\refstepcounter{equation}
\begin{gathered}
\;\;\; S_{\hspace{1pt}[x,\chi],[a,\rho]} \,\; = \\ 
\frac{1}{\text{dim}(\rho) \hspace{-1pt}\cdot \hspace{-1pt} |G|} \; \cdot \!\! \sum_{\substack{\; g \hspace{1pt} \in \hspace{1pt} G \hspace{1pt}: \\[1pt] {}^{g\hspace{-1pt}}x \hspace{1pt} \in \hspace{1pt} G_a}} \!\braket{\tau_x(\lambda)(g),a} \hspace{1pt}\cdot \hspace{1pt} \text{Tr}_{\rho}({}^{g\hspace{-0.7pt}}x) \hspace{1pt} \cdot \hspace{1pt} \chi(a^g) \; ,
\end{gathered}
\tag*{{\raisebox{-2.2em}{(\theequation)}}}
\end{equation}
where the index labels of the matrix are given by
\begin{enumerate}
\item pairs $(x,\chi)$ consisting of a representative of a conjugacy class $[x]\in  \text{Cl}(G)$ and a character $\chi \in (A_x)^{\vee}$ on the subgroup $A_x \subset A$ of $x$-invariant 1-form defects,
\item pairs $(a,\rho)$ consisting of a representative of a $G$-orbit $[a] \in A / G$ in $A$ and an irreducible representation $\rho$ of the stabiliser subgroup $G_a \subset G$ of $a$,
\end{enumerate}
and $\tau(\lambda)$ denotes the \textit{transgression} of a mixed 't Hooft anomaly\footnote{Here, we denote by $A^{\vee} := \text{Hom}(A,U(1))$ the \textit{Pontryagin dual group} of the finite abelian group $A$.} $[\lambda] \in H^2(G,A^{\vee})$ between $G$ and $A$.

\section{Two Dimensions}
\label{sec-two-dimensions}

In this section, we review the construction of the tube algebra $\text{Tube}(\mathcal{C})$ associated to a fusion category $\mathcal{C}$ based on \cite{Neshveyev2018,Izumi2000,Lin2023}. We describe its canonical $\ast$-structure and classify irreducible $\ast$-representations in several examples of invertible and non-invertible symmetries.

\subsection{Preliminaries}

In two dimensions, the finite (bosonic) generalised symmetries of a quantum field theory are described by a fusion category $\mathcal{C}$, whose objects $A,B \in \mathcal{C}$ correspond to topological line defects and whose morphisms $\varphi: A \to B$ correspond to topological junctions between lines:
\begin{equation}
\vspace{-5pt}
\begin{gathered}
\includegraphics[height=1.2cm]{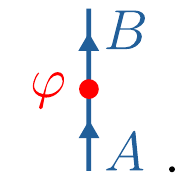}
\end{gathered}
\end{equation}
The monoidal structure $\otimes: \mathcal{C} \times \mathcal{C} \to \mathcal{C}$ captures the fusion of topological lines and their junctions:
\begin{equation}
\vspace{-5pt}
\begin{gathered}
\includegraphics[height=1.2cm]{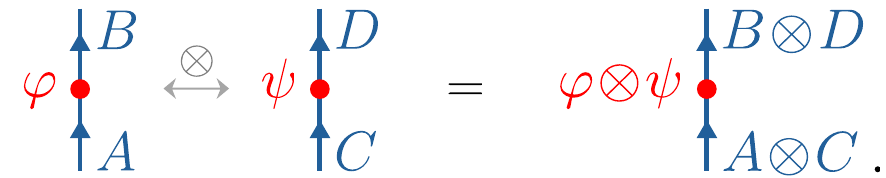}
\end{gathered}
\end{equation}
Moreover, we assume $\mathcal{C}$ to be equipped with the following additional structures:
\begin{enumerate}[label=\textbf{\arabic*}.]
\item \textbf{Dual structure:} A \textit{dual structure} on $\mathcal{C}$ allows us to bend topological line defects around in the sense that for each $A \in \mathcal{C}$ there exists a \textit{dual object} $A^{\vee} \in \mathcal{C}$ together with evaluation and coevaluation morphisms
\begin{equation}
\vspace{-5pt}
\begin{gathered}
\includegraphics[height=1.17cm]{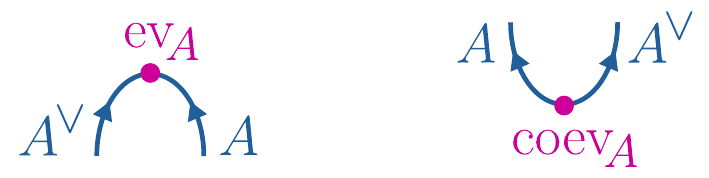}
\end{gathered}
\end{equation}
satisfying suitable \textit{zig-zag relations} \cite{Bakalov2001}. The assignment $A \mapsto A^{\vee}$ then extends to an op-monoidal functor\footnote{Here, $\mathcal{C}^{\text{op}}$ is the category with the same objects as $\mathcal{C}$ and morphisms given by $\text{Hom}_{(\mathcal{C}^{\text{op}})}(x,y) = \text{Hom}_{\mathcal{C}}(y,x)$ for all $x,y \in \mathcal{C}$.} $\vee: \mathcal{C} \to \mathcal{C}^{\text{op}}$. A \textit{pivotal structure} on $\mathcal{C}$ is a natural isomorphism $\xi: \vee^2 \Rightarrow \text{id}_{\mathcal{C}}$. The latter is called \textit{spherical} if the associated left and right traces of endomorphisms $\varphi \in \text{End}_{\mathcal{C}}(A)$ agree, i.e.
\begin{equation}
\vspace{-5pt}
\begin{gathered}
\;\;\quad\includegraphics[height=1.175cm]{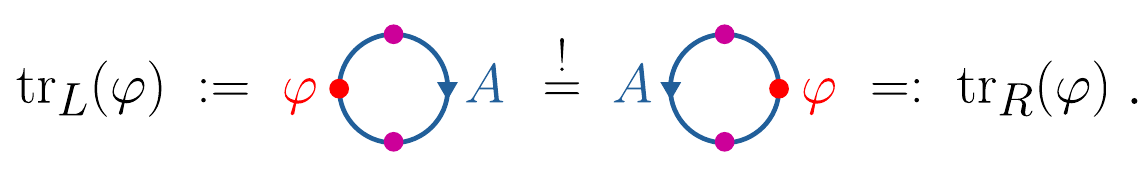}
\vspace{-8pt}
\end{gathered}
\vspace{8pt}
\end{equation}
In this case, we define the dimension of an object $A \in \mathcal{C}$ by $\text{dim}(A) := \text{tr}_{L/R}(\text{id}_A) \in \mathbb{C}$.

\item \textbf{Dagger structure:} A $\dagger$-structure on $\mathcal{C}$ allows us to reflect topological junctions in the sense that there exists a functor $\dagger: \mathcal{C} \to \mathcal{C}^{\text{op}}$ that acts as the identity on objects and anti-linearly on morphisms via
\begin{equation}
\vspace{-5pt}
\begin{gathered}
\includegraphics[height=1.2cm]{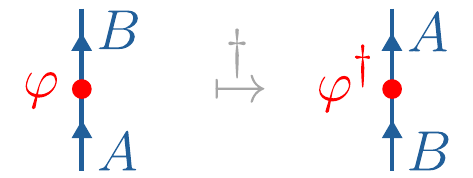}
\end{gathered}
\end{equation}
such that $\dagger^2 = \text{id}_{\mathcal{C}}$ and $\text{End}_{\mathcal{C}}(A)$ is a C*-algebra for all $A \in \mathcal{C}$. We assume this $\dagger$-structure to be compatible with the monoidal structure on $\mathcal{C}$ in the sense that $\otimes: \mathcal{C} \times \mathcal{C} \to \mathcal{C}$ is a $\dagger$-functor\footnote{A functor $F: \mathcal{C} \to \mathcal{C}'$ between two $\dagger$-categories $\mathcal{C}$ and $\mathcal{C}'$ is called a \textit{$\dagger$-functor} if $\dagger' \circ F = F^{\text{op}} \circ \dagger$.} with unitary coherence data\footnote{We collectively refer to any natural coherence isomorphisms as \textit{coherence data} in what follows. In the above case, the coherence data associated to the product $\otimes$ is given by the associator and left and right unitor isomorphisms.}. In this case, we call $\mathcal{C}$ a \textit{unitary fusion category} \cite{Penneys2021}.
\end{enumerate}
We assume the above structures to be compatible with one another in the sense that $\vee: \mathcal{C} \to \mathcal{C}^{\text{op}}$ is a $\dagger$-functor with unitary coherence data. In this case,
\begin{equation}
\label{eq-unitary-pivotal}
\vspace{-5pt}
\begin{gathered}
\includegraphics[height=1.6cm]{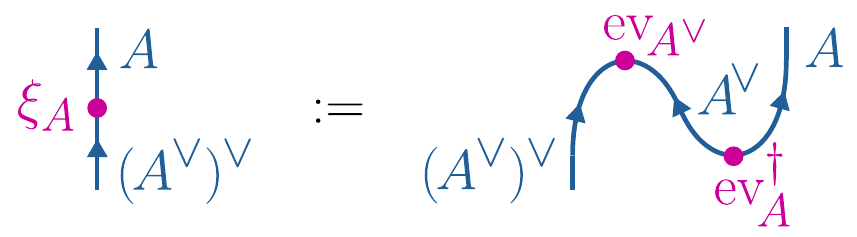}
\end{gathered}
\end{equation}
defines a canonical pivotal structure on $\mathcal{C}$. A dual structure of this type is called a \textit{unitary dual structure}. There exists a unique unitary dual structure $\vee$ on $\mathcal{C}$ such that (\ref{eq-unitary-pivotal}) is spherical \cite{Penneys2018}. In the following, we assume $\mathcal{C}$ to be a unitary fusion category equipped with the unique unitary dual structure such that the associated canonical pivotal structure (\ref{eq-unitary-pivotal}) is spherical.

\subsection{Tube Algebra}

Given the symmetry category $\mathcal{C}$, we can associate to it its \textit{tube algebra} $\text{Tube}(\mathcal{C})$, which captures the linking action of topological line defects on twisted sector local operators in two dimensions. As a vector space, the tube algebra decomposes into a direct sum \cite{Neshveyev2018,Bhowmick2018}
\begin{equation}
\text{Tube}(\mathcal{C}) \;\;\; =  \bigoplus_{[X],\,[Y] \, \in \, \pi_0(\mathcal{C})} \! \text{Tube}(\mathcal{C})_{X,Y} \; ,
\end{equation}
where the summand associated to simple lines $X,Y \in \mathcal{C}$ is given by the quotient space
\begin{equation}
\text{Tube}(\mathcal{C})_{X,Y} \; = \; \bigoplus_{A \hspace{1pt} \in \hspace{1pt} \mathcal{C}} \, \text{Hom}_{\mathcal{C}}(A \otimes X \hspace{1pt}, Y \otimes A) \; \Big/ \, \sim
\end{equation}
of local intersection morphisms
\begin{equation}
\label{eq-2d-tube-objects}
\vspace{-5pt}
\begin{gathered}
\includegraphics[height=1.2cm]{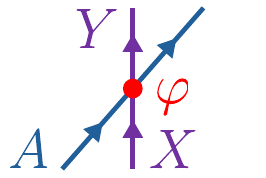}
\end{gathered}
\end{equation} 
subjected to the equivalence relation generated by
\begin{equation}
\vspace{-5pt}
\begin{gathered}
\includegraphics[height=1.66cm]{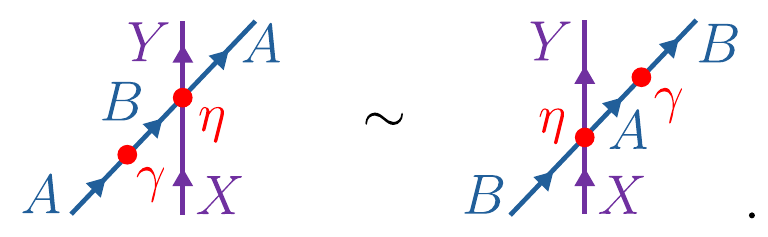}
\end{gathered}
\end{equation}
We denote the equivalence class of morphisms (\ref{eq-2d-tube-objects}) by 
\begin{equation}
\tub[\hspace{-2.4pt}X]{\hspace{0.5pt}Y}{\hspace{-1pt}A}{\scriptstyle\hspace{1pt} \varphi} \; \in \; \text{Tube}(\mathcal{C})_{X,Y} \; .
\end{equation}
As a vector space, $\text{Tube}(\mathcal{C})_{X,Y}$ is isomorphic to
\begin{equation}
\text{Tube}(\mathcal{C})_{X,Y} \;\; \cong  \bigoplus_{[A] \hspace{1pt} \in \hspace{1pt}  \pi_0(\mathcal{C})} \text{Hom}_{\mathcal{C}}\big(A \otimes X, Y \otimes A\big) \, ,
\end{equation}
which shows that the tube algebra is finite-dimensional. The algebra multiplication is induced by
\begin{equation}
\label{eq-2d-tube-multi}
\vspace{-5pt}
\begin{gathered}
\includegraphics[height=1.95cm]{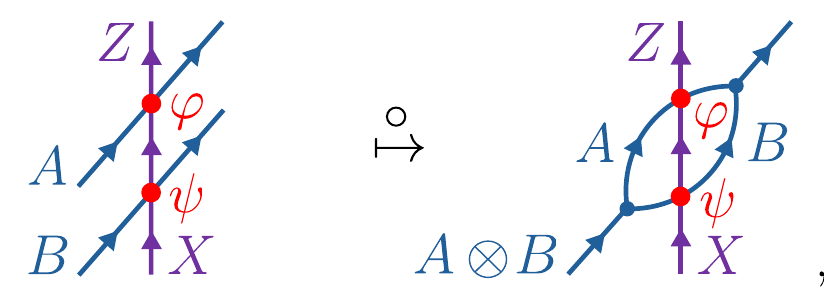}
\end{gathered}
\end{equation}
which we denote by
\begin{equation}
\tub[\hspace{-3.2pt}Y^{\hspace{-0.5pt}\prime}]{\hspace{0.5pt}Z}{\hspace{-1.5pt}A}{\hspace{1pt} \varphi} \; \circ \; \tub[\hspace{-2.4pt}X]{\hspace{0.5pt}Y}{\hspace{-1pt}B}{\raisebox{-0.4pt}{\hspace{1pt}$\scriptstyle \psi$}} \;\; = \;\; \delta_{\hspace{1.2pt}Y,Y'} \, \cdot \, \tuub[\hspace{-2.4pt}X]{\hspace{0.4pt}Z}{\raisebox{0.7pt}{\hspace{-10pt}$\scriptstyle A \hspace{0.5pt} \otimes B$}}{\hspace{1pt} \varphi \hspace{0.6pt}\circ\hspace{0.6pt} \psi}
\end{equation}
on the generators of $\text{Tube}(\mathcal{C})$. The tube algebra further admits an antilinear involution $\ast: \text{Tube}(\mathcal{C}) \to \text{Tube}(\mathcal{C})$, which is induced by \cite{Izumi2000}
\begin{equation}
\label{eq-2d-tube-invol}
\vspace{-5pt}
\begin{gathered}
\includegraphics[height=1.37cm]{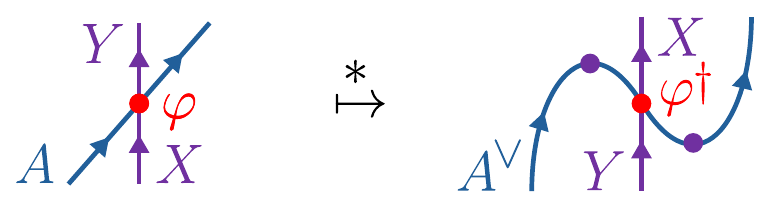}
\end{gathered}
\end{equation}
and which we denote by
\begin{equation}
\tub[\hspace{-2.4pt}X]{\hspace{0.5pt}Y}{\hspace{-1pt}A}{\hspace{1pt} \varphi}^{\ast} \; = \; \tub[\hspace{-1.2pt}Y]{\hspace{0pt}X}{\hspace{-2pt}A^{\hspace{-0.8pt}\vee}}{\raisebox{-2.2pt}{\hspace{1pt}$\scriptstyle \varphi^{\dagger}$}}
\end{equation}
on the generators of the tube algebra.

\subsection{Examples}

We conclude this section with examples of unitary fusion category symmetries and $\ast$-representations of their associated tube algebras. We discuss anomalous invertible group-like symmetries as well as non-invertible symmetries of Tambara-Yamagami and Fibonacci type.

\subsubsection{Group Symmetry}

We first consider the symmetry category $\mathcal{C} = \text{Hilb}_G^{\omega}$ corresponding to a finite group $G$ with 't Hooft anomaly $[\omega] \in H^3(G,U(1))$. Simple objects in $\mathcal{C}$ are given by group elements $g \in G$ that fuse according to the group law of $G$ with associator
\begin{equation}
\vspace{-5pt}
\begin{gathered}
\includegraphics[height=1cm]{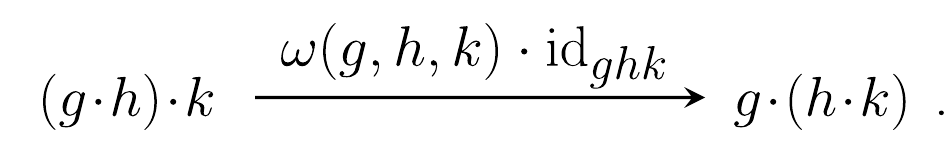}
\end{gathered}
\end{equation}
The duals and dimensions of simple objects are given by $g^{\vee} = g^{-1}$ and $\text{dim}(g)=1$, respectively. This symmetry category is unitary for all $G$ and $\omega$.

\textbf{Tube algebra:} As discussed in \cite{Bullivant2019,Bartsch2023a}, the tube algebra of $\mathcal{C} = \text{Hilb}_G^{\omega}$ is the twisted Drinfeld double \cite{Drinfeld1989,Majid1991DoublesOQ,Roche1990}
\begin{equation}
\text{Tube}(\text{Hilb}_G^{\omega}) \; = \; \mathcal{D}^{\hspace{1pt}\omega}(G) \; .
\end{equation}
The latter has generators\footnote{Here, we use the notation ${}^gx:=gxg^{-1}$ for the conjugation action of a group element $g\in G$ on a group element $x \in G$.} $\tub[x]{\hspace{-2.7pt}{}^{\mathrlap{\raisebox{-2.5pt}{\hspace{-5pt}\crule[white]{8pt}{8pt}}}g}\hspace{-0.95pt}x}{\raisebox{1pt}{$\scriptstyle g$}}{} \; (g,x \in G)$, whose algebra multiplication is given by
\begin{equation}\label{eq-grp-alg-mult}
\tub[x]{\hspace{-2.7pt}{}^{\mathrlap{\raisebox{-2.5pt}{\hspace{-5pt}\crule[white]{8pt}{8pt}}}g}\hspace{-0.95pt}x}{\raisebox{1pt}{$\scriptstyle g$}}{} \, \circ \, \tub[\raisebox{1pt}{$\scriptstyle y$}]{\raisebox{1pt}{$\scriptstyle \hspace{-2.7pt}{}^{\mathrlap{\raisebox{-1.5pt}{\hspace{-5pt}\crule[white]{8pt}{8pt}}}h}\hspace{-0.95pt}y$}}{h}{} \,\; = \;\, \delta_{x,{}^hy} \hspace{1pt} \cdot \hspace{1pt} \tau_y(g,h) \hspace{1pt} \cdot \hspace{1pt} \tub[\raisebox{1pt}{$\scriptstyle y$}]{\raisebox{1pt}{$\scriptstyle \hspace{-6.7pt}{}^{\mathrlap{\raisebox{-2.5pt}{\crule[white]{6pt}{6pt}}} g\hspace{-0.5pt}h}\hspace{-0.8pt} y$}}{\hspace{-2.2pt} gh}{}  \; .
\end{equation}
Here, we defined the multiplicative phase
\begin{equation}
\tau_x(\omega)(g,h) \; := \; \frac{\omega(g,h,x) \cdot \omega({}^{gh}x,g,h)}{\omega(g,{}^hx,h)}
\end{equation}
known as the \textit{transgression} of the 't Hooft anomaly $\omega$. As a result of the cocycle condition for $\omega$, it satisfies
\begin{equation}
\label{eq-transgression-cocycle-cond}
(d\tau)_x(g,h,k) \;\, := \;\, \frac{\tau_x(h,k) \cdot \tau_x(g,hk)}{\tau_x(gh,k) \cdot \tau_{({}^kx)}(g,h)} \,\; = \;\, 1 \, ,
\end{equation}
which ensures that the algebra multiplication (\ref{eq-grp-alg-mult}) is associative. Using (\ref{eq-2d-tube-invol}), the $\ast$-structure on $\mathcal{D}^{\hspace{1pt}\omega}(G)$ can be computed to be
\begin{equation}
\label{eq-drinfeld-double-invol}
\tub[x]{\hspace{-2.7pt}{}^{\mathrlap{\raisebox{-2.5pt}{\hspace{-5pt}\crule[white]{8pt}{8pt}}}g}\hspace{-0.95pt}x}{\raisebox{1pt}{$\scriptstyle g$}}{}^{\ast} \; = \; \mu_x(g) \hspace{1pt} \cdot \hspace{1pt} \tub[\hspace{-3pt}{}^g\hspace{-0.95pt}x]{x}{\raisebox{1pt}{$\scriptstyle g^{-1}$}}{} \; ,
\end{equation}
where we defined the multiplicative phase
\begin{equation}
\mu_x(g) \; := \; \tau^{\ast}_x(g^{-1},g) \; .
\end{equation}
As a consequence of (\ref{eq-transgression-cocycle-cond}), it satisfies
\begin{equation}
\mu_x(g^{-1}) \; = \; \mu_{(x^g)}(g) \qquad \text{and} \qquad d\mu \; = \; \widehat{\tau} \hspace{1pt}/\hspace{1pt} \tau \; ,
\end{equation}
where we defined the 2-cocycle
\begin{equation}
\widehat{\tau}_x(g,h) \; := \; \tau^{\ast}_{({}^{gh}x)}(h^{-1},g^{-1}) \; .
\end{equation}
This ensures that the $\ast$-structure (\ref{eq-drinfeld-double-invol}) is involutory and compatible with the algebra multiplication.

\textbf{Tube representations:} In order to classify irreducible $\ast$-representations $R$ of $\mathcal{D}^{\hspace{1pt}\omega}(G)$, we use the equivalence (\ref{eq-classif-star-reps}) as well as the fact that \cite{Davydov2017,Gruen2021}
\begin{equation}
\mathcal{Z}(\text{Hilb}_G^{\omega}) \;\;\; = \bigoplus_{[x] \, \in \, \text{Cl}(G)} \text{Rep}^{\tau_x(\omega)}(G_x) \, ,
\end{equation}
which shows that we can label each such $R$ by a pair $(x,\rho)$ consisting of 
\begin{enumerate}
\item a representative $x$ of a conjugacy class $[x] \in \text{Cl}(G)$,
\item an irreducible representation $\rho$ of the centraliser $G_x$ of $x$ with projective 2-cocycle $\tau_x(\omega) \in Z^2(G_x,U(1))$.
\end{enumerate}
The associated irreducible representation $R = R_{(x,\rho)}$ of $\mathcal{D}^{\hspace{1pt}\omega}(G)$ can be constructed via induction \cite{Willerton2008}: To this end, we fix for each $y \in [x]$ a representative $r_y \in G$ such that ${}^{(r_y)}y = x$ (with $r_x := e$). Using these, we can define
\begin{equation}
g_y \; := \; r_{({}^gy)} \cdot g \cdot r_y^{-1} \; \in \; G_x
\end{equation}
for all $g \in G$ and $y \in [x]$. If we denote by $\mathcal{V}$ the Hilbert space underlying the (projective) representation $\rho$ of $G_x$, then $R_{(x,\rho)}$ acts on the Hilbert space $\mathcal{H}$ with $G$-grading
\begin{equation}
\mathcal{H}_y \; = \; \begin{cases} \mathcal{V} &\text{if} \; y \in [x] \\ 0 &\text{otherwise} \end{cases}
\end{equation}
via the induced tube action
\begin{equation}
R_{(x,\rho)}\Big( \tub[\raisebox{1pt}{$\scriptstyle y$}]{\raisebox{1pt}{$\scriptstyle \hspace{-2.7pt}{}^{\mathrlap{\raisebox{-2.5pt}{\hspace{-1pt}\crule[white]{4pt}{8pt}}}g}\hspace{-0.95pt}y$}}{\raisebox{1pt}{$\scriptstyle g$}}{} \Big) \; := \; \kappa_y(g) \cdot \rho(g_y ) \; ,
\end{equation}
where we defined the multiplicative phase
\begin{equation}
\kappa_y(g) \; := \; \frac{\tau_y(r_{({}^gy)},g)}{\tau_y(g_y,r_y)} \; .
\end{equation}
As a consequence of (\ref{eq-transgression-cocycle-cond}), it satisfies
\begin{equation}
\frac{\kappa_y(h) \cdot \kappa_{({}^hy)}(g)}{\kappa_y(gh)} \; = \; \frac{\tau_y(g,h)}{\tau_x( g_{({}^hy)}, h_y)} \; ,
\end{equation}
which ensures that $R_{(x,\rho)}$ respects the algebra multiplication of $\mathcal{D}^{\hspace{1pt}\omega}(G)$. It is straightforward to check that $R_{(x,\rho)}$ is a $\ast$-representation of $\mathcal{D}^{\hspace{1pt}\omega}(G)$ if and only if $\rho$ is a unitary projective representation of $G_x$. Since every finite-dimensional representation of a finite group is equivalent to a unitary one, this shows that all irreducible representations of $\mathcal{D}^{\hspace{1pt}\omega}(G)$ are $\ast$-representations as expected.

\subsubsection{Tambara-Yamagami Symmetry}

As a second example, let us consider a symmetry category $\mathcal{C} = \text{TY}_A^{\chi,s}$ of Tambara-Yamagami type \cite{Tambara1998}, which is specified by the following pieces of data:
\begin{enumerate}
\item A finite abelian group $A$,
\item a non-deg. symmetric bicharacter $\chi: A \times A \to U(1)$,
\item a square-root $s$ of $1/|A|$.
\end{enumerate}
The simple objects of $\mathcal{C}$ consist of group elements $a \in A$ together with a non-invertible defect $m$, which are subject to the fusion rules
\begin{equation}
\refstepcounter{equation}
a \otimes b \hspace{1pt} = \hspace{1pt} a \cdot b \hspace{1pt} , \quad a \otimes m \hspace{1pt} = \hspace{1pt} m \otimes a = m \hspace{1pt} , \quad m \otimes m \hspace{1pt} =  \hspace{1pt} \bigoplus_{a \hspace{1pt} \in \hspace{1pt} A} a \hspace{1pt} .
\tag*{{\raisebox{-1.5em}{(\theequation)}}}
\end{equation}
The non-trivial components of the associator are
\begin{equation}
\vspace{-5pt}
\begin{gathered}
\includegraphics[height=4.8cm]{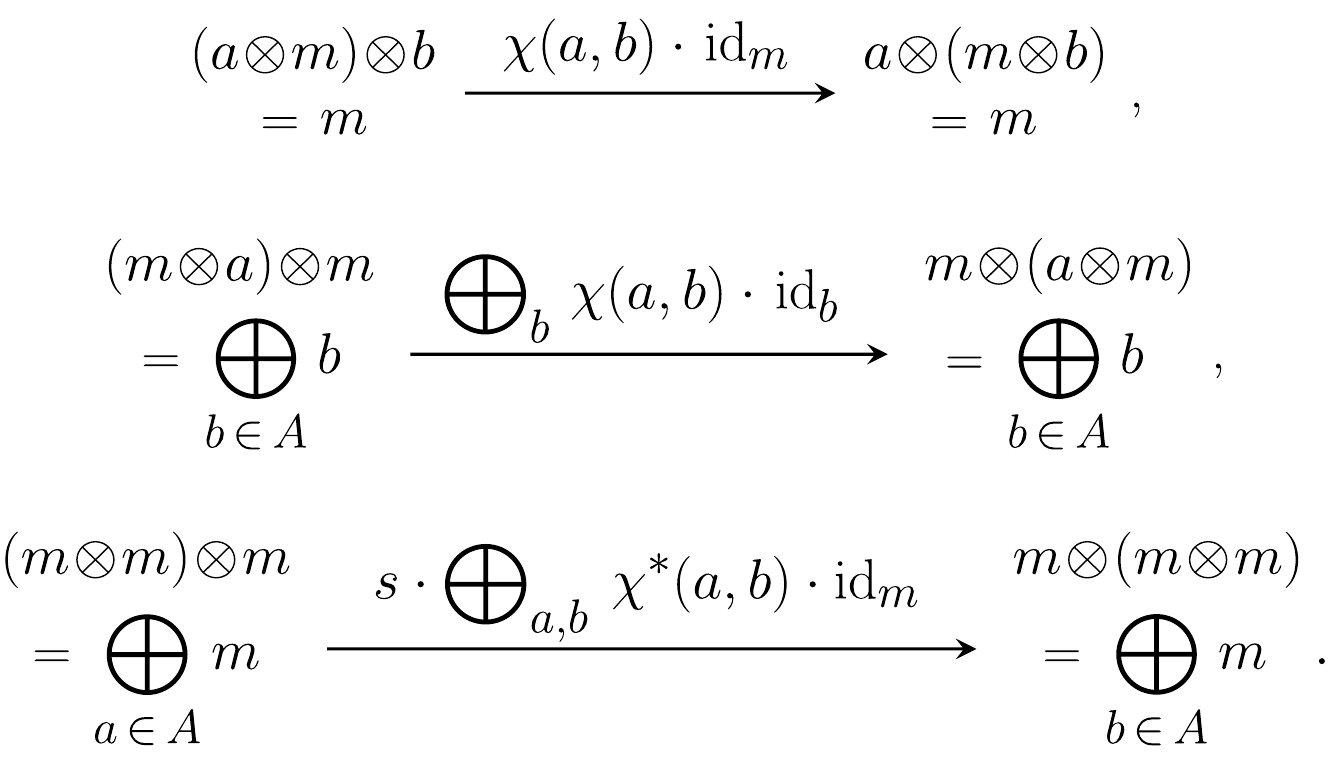}
\vspace{-2pt}
\end{gathered}
\vspace{10pt}
\end{equation}
The duals of simple objects are $a^{\vee} = a^{-1}$ and $m^{\vee} =m$. Their dimensions are $\text{dim}(a) = 1$ and $\text{dim}(m) = \sqrt{|A|}$. This symmetry category is unitary for all $A$, $\chi$ and $s$.

\textbf{Tube algebra:} The tube algebra of $\mathcal{C} = \text{TY}_A^{\chi,s}$ 
has the following types of generators:
\begin{equation}
\tub[x]{x}{a}{} \, , \quad \tub[x]{y}{\!m}{} \, , \quad \tub[\hspace{-2.5pt} m]{m}{\hspace{-2pt}m}{a} \, , \quad \tub[\hspace{-2.5pt} m]{m}{a}{} \, ,
\end{equation}
where $a,x,y \in A$. Using (\ref{eq-2d-tube-multi}), their algebra multiplication can be determined to be
\begin{equation}
\begin{aligned}
\tub[x]{x}{a}{} \, \circ \, \tub[x]{x}{b}{} \; &= \; \tub[x]{x}{\hspace{-2.2pt} ab}{} \, , \\[3pt]
\tub[x]{y}{\!m}{} \, \circ \, \tub[x]{x}{a}{} \; &= \; \chi(a,y) \cdot \tub[x]{y}{\!m}{} \, , \\[3pt]
\tub[y]{y}{a}{} \, \circ \, \tub[x]{y}{\!m}{} \; &= \; \chi(a,x) \cdot \tub[x]{y}{\!m}{} \, , \\[3pt]
\tub[y]{z}{\!m}{} \, \circ \, \tub[x]{y}{\!m}{} \; &= \; \delta_{x,z} \cdot \chi^{\ast}(x,y) \\[3pt]
&\hphantom{= \;\,} \cdot \hspace{1pt}\sum\nolimits_a \chi^{\ast}(a,y) \cdot \tub[x]{x}{a}{} \, , \\[4pt]
\tub[\hspace{-2.5pt} m]{m}{a}{} \, \circ \, \tub[\hspace{-2.5pt} m]{m}{b}{} \; &= \; \chi^{\ast}(a,b) \cdot \tub[\hspace{-2.5pt} m]{m}{\hspace{-2.2pt} ab}{} \, , \\[3pt]
\tub[\hspace{-2.5pt} m]{m}{a}{} \, \circ \, \tub[\hspace{-2.5pt} m]{m}{\hspace{-2pt}m}{b} \; &= \; \tub[\hspace{-2.5pt} m]{m}{\hspace{-2pt}m}{b} \, \circ \, \tub[\hspace{-2.5pt} m]{m}{a}{} \\[4pt] 
&= \; \chi(a,ab) \cdot \tub[\hspace{-2.5pt} m]{m}{\hspace{-2pt}m}{ab} \, , \\[3pt]
\tub[\hspace{-2.5pt} m]{m}{\hspace{-2pt}m}{a} \, \circ \, \tub[\hspace{-2.5pt} m]{m}{\hspace{-2pt}m}{b} \; &= \; \frac{s}{|A|} \cdot \chi(a,b) \\[3pt]
&\hphantom{= \;\,}\cdot \, \sum\nolimits_c \chi^{\ast}(ab,c) \cdot \tub[\hspace{-2.5pt} m]{m}{c}{} \, .
\end{aligned}
\end{equation}
Using (\ref{eq-2d-tube-invol}), the $\ast$-structure can be computed to be 
\begin{equation}
\begin{aligned}
\tub[x]{x}{a}{}^{\ast} \; &= \; \tub[x]{x}{\! a^{\scalebox{0.4}[0.4]{$-1$}}}{} \; , \\[3pt]
\tub[x]{y}{\!m}{}^{\ast} \; &= \; \chi(x,y) \cdot \tub[y]{x}{\!m}{} \; , \\[3pt]
\tub[\hspace{-2.5pt} m]{m}{a}{}^{\ast} \; &= \; \chi^{\ast}(a,a) \cdot \tub[\hspace{-2.5pt} m]{m}{\! a^{\scalebox{0.4}[0.4]{$-1$}}}{} \; , \\[3pt]
\tub[\hspace{-2.5pt} m]{m}{\hspace{-2pt}m}{a}^{\ast} \; &= \; s \cdot \sum_b \, \chi^{\ast}(a,b) \cdot \tub[\hspace{-2.5pt} m]{m}{\hspace{-2pt}m}{b} \; .
\end{aligned}
\end{equation}
Associativity of the algebra multiplication as well as its compatibility with the $\ast$-structure then holds as a consequence of the character identity 
\begin{equation}
\frac{1}{|A|} \, \cdot \sum_{a \, \in \, A} \chi(a,b) \,\; = \;\, \delta_{b,e} \; .
\end{equation}

\textbf{Tube representations:} Using the equivalence (\ref{eq-classif-star-reps}), we can find the irreducible $\ast$-representations of $\text{Tube}(\text{TY}_A^{\chi,s})$ from the classification of simple objects in the Drinfeld center of $\text{TY}_A^{\chi,s}$ \cite{Gelaki2009}. Concretely, there are $\frac{1}{2} \hspace{1pt}|A| \!\cdot\!(|A| + 7)$ irreducible representations of $\text{Tube}(\text{TY}_A^{\chi,s})$, which can be grouped into the following three categories:
\begin{itemize}
\item There are $2 \! \cdot \! |A|$ one-dimensional representations $R^{\pm}_x$ labelled by elements $x \in A$, which act on the non-trivial twisted sector $\mathcal{H}_x \cong \mathbb{C}$ via
\begin{equation}
\begin{aligned}
R^{\pm}_x\Big( \tub[x]{x}{a}{} \Big) \; &= \; \chi(a,x) \, , \\[3pt]
R^{\pm}_x\Big( \tub[x]{x}{\!m}{} \Big) \; &= \; \frac{1}{s} \cdot (\pm\Delta_x) \, ,
\end{aligned}
\end{equation}
where $\Delta_x$ is a square-root of $\chi^{\ast}(x,x) \in U(1)$.

\item There are $2 \! \cdot \! |A|$ one-dimensional representations $R^{\pm}_{\rho}$ labelled by antiderivatives\footnote{An antiderivative of $\chi$ is a map $\rho: A \to U(1)$ such that $(d\rho)(a,b) := \rho(a)\cdot \rho(b) \cdot \rho^{\ast}(ab) \equiv \chi(a,b)$ for all $a,b \in A$. Since $\chi$ is symmetric, such a $\rho$ always exists and the set of all antiderivatives forms a torsor over $A^{\vee} = \text{Hom}(A,U(1))$. This shows that there are $|A^{\vee}| = |A|$ antiderivatives of $\chi$.} $\rho$ of $\chi$, which act on the non-trivial twisted sector $\mathcal{H}_m \cong \mathbb{C}$ via
\begin{equation}
\begin{aligned}
R^{\pm}_{\rho}\Big( \tub[\hspace{-2.5pt} m]{m}{a}{} \Big) \; &= \; \rho^{\ast}(a) \, , \\[3pt]
R^{\pm}_{\rho}\Big( \tub[\hspace{-2.5pt} m]{m}{\hspace{-2pt}m}{a} \Big) \; &= \; s \cdot (\pm \Delta_{\rho}) \cdot \rho(a^{-1}) \, ,
\end{aligned}
\end{equation}
where $\Delta_{\rho}$ is a aquare-root of $s \cdot \sum_c \rho^{\ast}(c) \in U(1)$.

\item There are $\frac{1}{2}|A|\!\cdot \!(|A|-1)$ two-dimensional representations $R_{x,y}$ labelled by distinct elements $x,y \in A$, which act on the twisted sectors $\mathcal{H}_x \cong \mathcal{H}_y \cong \mathbb{C}$ via
\begin{equation}
\begin{aligned}
R_{x,y}\Big( \tub[x]{x}{a}{} \Big) \; &= \; \begin{pmatrix}
\chi(a,y) & \;0\; \\
0 & \;0\;
\end{pmatrix} \, , \\[3pt]
R_{x,y}\Big( \tub[y]{y}{a}{} \Big) \; &= \; \begin{pmatrix}
\;0\; & 0 \\
\;0\; & \chi(a,x)
\end{pmatrix} \, , \\[3pt]
R_{x,y}\Big( \tub[x]{y}{\!m}{} \Big) \; &= \; \frac{1}{s} \cdot \begin{pmatrix}
0 & \;0\; \\
\chi^{\ast}(x,y) & \;0\;
\end{pmatrix} \, , \\[3pt]
R_{x,y}\Big( \tub[y]{x}{\!m}{} \Big) \; &= \; \frac{1}{s} \cdot 
\begin{pmatrix}
\,0\; & \;1\, \\
\,0\; & \;0\,
\end{pmatrix} \, .
\end{aligned}
\end{equation}
\end{itemize}
It is easy to check that all of the above are $\ast$-represen-tations as expected. In the case $A = \mathbb{Z}_2$ and $s = 1/\sqrt{2}$, this reproduces the Ising symmetry category and its known action on twisted sector local operators in the two-dimensional critical Ising CFT \cite{Chang2019,Thorngren2024}.

\subsubsection{Fibonacci Symmetry}

As a last example, we consider a symmetry category $\mathcal{C}$ with only two simple objects, $1$ and $\tau$, whose fusion rules are given by
\begin{equation}
\label{eq-fib-fusion-rules}
\tau \otimes \tau \; = \; 1 \oplus \tau \, .
\end{equation}
The pentagon equation then admits the following solution for the associator \cite{Davydov2011}, 
\begin{equation}
\vspace{-5pt}
\begin{gathered}
\includegraphics[height=1.47cm]{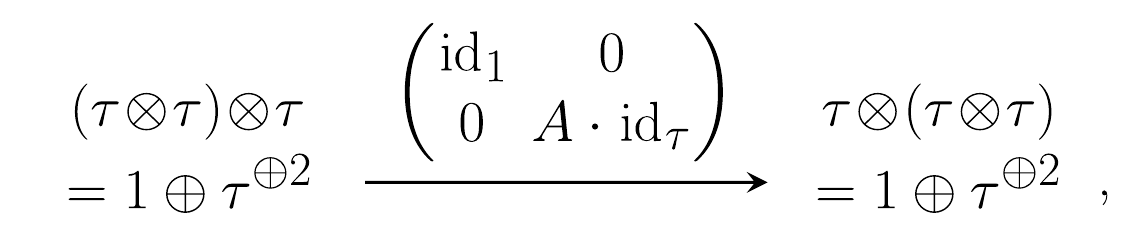}
\end{gathered}
\end{equation}
where the self-inverse $(2 \!\times \!2)$-matrix $A$ is given by
\begin{equation}
A \; = \;
\begin{pmatrix}
-a \; & 1/\lambda \\
-a\lambda \; & a
\end{pmatrix} \; .
\end{equation}
Here, $a \in \mathbb{R}$ is one of the two solutions of the quadratic equation $a^2 = a + 1$ given by
\begin{equation}
a_{\pm} \; = \; \frac{1}{2} \, \big(1 \pm \sqrt{5} \big) \, ,
\end{equation}
where $a_+ \equiv - 1/a_- \approx 1.62$ is the golden ratio. Furthermore, $\lambda \in \mathbb{C}^{\times}$ is a gauge parameter that describes a family of equivalent fusion categories for fixed $a$. Up to equivalence, there are hence only two distinct fusion categories with fusion rules (\ref{eq-fib-fusion-rules}) -- the \textit{Fibonacci} and the \textit{Yang-Lee category} -- which we denote by $\text{Fib}^+$ and $\text{Fib}^{-}$ and which correspond to choosing $a=a_-$ and $a = a_+$, respectively. The reason for this (perhaps confusing) notation is that the dimension of the non-invertible self-dual object $\tau$ in each case is given by
\begin{equation}
\text{dim}(\tau) \; = \; - \, \frac{1}{a} \; \equiv \;
\begin{cases}
a_+ > 0 &\text{for} \; \text{Fib}^+ \\ a_- < 0 &\text{for} \; \text{Fib}^-
\end{cases}
\; .
\end{equation}
Since the matrix $A$ is unitary if and only if
\begin{equation}
|\lambda|^2 \; = \; - \, \frac{1}{a} \; ,
\end{equation}
we see that only $\text{Fib}^+$ can be made unitary, whereas $\text{Fib}^-$ represents its non-unitary counterpart.

\textbf{Tube algebra:} The tube algebra associated to $\text{Fib}^{\pm}$ is spanned by seven generators
\begin{equation}
\begin{gathered}
\tub[1]{1}{1}{} \, , \quad \tub[1]{1}{\tau}{} \, , \quad \tub[1]{\tau}{\tau}{} \, , \quad \tub[\tau]{1}{\tau}{} \, , \\[5pt] 
\quad \tub[\tau]{\tau}{1}{} \, , \quad \tub[\tau]{\tau}{\tau}{\raisebox{-1.5pt}{$\scriptstyle 1$}} \, , \quad \tub[\tau]{\tau}{\tau}{\tau} \, ,
\end{gathered}
\end{equation}
whose algebra multiplication can be computed to be
\begin{equation}
\refstepcounter{equation}
\begin{aligned}
\tub[1]{1}{\tau}{} \, \circ \, \tub[1]{1}{\tau}{} \; &= \; \tub[1]{1}{1}{} \, + \, \tub[1]{1}{\tau}{} \; , \\[3pt]
\tub[1]{\tau}{\tau}{} \, \circ \, \tub[1]{1}{\tau}{} \; &= \; a \cdot \tub[1]{\tau}{\tau}{} \; , \\[3pt]
\tub[\tau]{1}{\tau}{} \, \circ \, \tub[1]{\tau}{\tau}{} \; &= \; \tub[1]{1}{1}{} \, + \, a \cdot \tub[1]{1}{\tau}{} \; , \\[3pt]
\tub[\tau]{\tau}{\tau}{\raisebox{-1.5pt}{$\scriptstyle 1$}} \, \circ \, \tub[1]{\tau}{\tau}{} \; &= \; -a \cdot \tub[1]{\tau}{\tau}{} \; , \\[3pt]
\tub[\tau]{\tau}{\tau}{\tau} \, \circ \, \tub[1]{\tau}{\tau}{} \; &= \; a^2 \cdot \tub[1]{\tau}{\tau}{} \; , \\[3pt]
\tub[1]{1}{\tau}{} \, \circ \, \tub[\tau]{1}{\tau}{} \; &= \; a \cdot \tub[\tau]{1}{\tau}{} \; , \\[3pt]
\tub[1]{\tau}{\tau}{} \, \circ \, \tub[\tau]{1}{\tau}{} \; &= \; -a \cdot \tub[\tau]{\tau}{1}{} \, + \, \tub[\tau]{\tau}{\tau}{\raisebox{-1.5pt}{$\scriptstyle 1$}} \\[3pt] &\quad\;\, + \, a^2 \cdot \tub[\tau]{\tau}{\tau}{\tau} \; , \\[3pt]
\tub[\tau]{1}{\tau}{} \, \circ \, \tub[\tau]{\tau}{\tau}{\raisebox{-1.5pt}{$\scriptstyle 1$}} \; &= \; -a \cdot \tub[\tau]{1}{\tau}{} \; , \\[3pt]
\tub[\tau]{1}{\tau}{} \, \circ \, \tub[\tau]{\tau}{\tau}{\tau} \; &= \; a^2 \cdot \tub[\tau]{1}{\tau}{} \; , \\[3pt]
\tub[\tau]{\tau}{\tau}{\raisebox{-1.5pt}{$\scriptstyle 1$}} \, \circ \, \tub[\tau]{\tau}{\tau}{\raisebox{-1.5pt}{$\scriptstyle 1$}} \; &= \; -a^3 \cdot \tub[\tau]{\tau}{1}{} \, + \, a^2 \cdot \tub[\tau]{\tau}{\tau}{\tau} \; , \\[3pt]
\tub[\tau]{\tau}{\tau}{\tau} \, \circ \, \tub[\tau]{\tau}{\tau}{\tau} \; &= \; -a^2 \cdot \tub[\tau]{\tau}{1}{}  \, + \, \tub[\tau]{\tau}{\tau}{\raisebox{-1.5pt}{$\scriptstyle 1$}}  \\[3pt] &\quad\;\, + \, a^3 \cdot \tub[\tau]{\tau}{\tau}{\tau} \; , \\[3pt]
\tub[\tau]{\tau}{\tau}{\raisebox{-1.5pt}{$\scriptstyle 1$}} \, \circ \, \tub[\tau]{\tau}{\tau}{\tau} \; &= \; \tub[\tau]{\tau}{\tau}{\tau} \, \circ \, \tub[\tau]{\tau}{\tau}{\raisebox{-1.5pt}{$\scriptstyle 1$}} \\[3pt] &= \; a^2 \cdot \tub[\tau]{\tau}{1}{} \, - \, a^2 \cdot \tub[\tau]{\tau}{\tau}{\tau} \; .
\end{aligned}
\tag*{{\raisebox{-1.4em}{(\theequation)}}}
\vspace{6pt}
\end{equation}
Here, $a=a_-$ for $\text{Fib}^+$ and $a = a_+$ for $\text{Fib}^-$. Using (\ref{eq-2d-tube-invol}), the $\ast$-structure\footnote{Although $\text{Fib}^-$ is not unitary, (\ref{eq-fib-tube-invol}) still defines an antilinear involution on $\text{Tube}(\text{Fib}^-)$. However, this $\ast$-structure is no longer positive w.r.t. the canonical linear functional $F$ as in (\ref{eq-positive-functional}).} can be computed to be 
\begin{equation}
\label{eq-fib-tube-invol}
\begin{aligned}
\tub[1]{1}{\tau}{}^{\ast} \; &= \; \tub[1]{1}{\tau}{} \; , \\[3pt]
\tub[1]{\tau}{\tau}{}^{\ast} \; &= \; -\frac{1}{a} \cdot \tub[\tau]{1}{\tau}{} \; , \\[3pt]
\tub[\tau]{1}{\tau}{}^{\ast} \; &= \; - \,a \cdot \tub[1]{\tau}{\tau}{} \; , \\[3pt]
\tub[\tau]{\tau}{\tau}{\raisebox{-1.5pt}{$\scriptstyle 1$}}^{\ast} \; &= \; -\,a \cdot \tub[\tau]{\tau}{\tau}{\raisebox{-1.5pt}{$\scriptstyle 1$}} \, - \, a\cdot \tub[\tau]{\tau}{\tau}{\tau} \; , \\[3pt]
\tub[\tau]{\tau}{\tau}{\tau}^{\ast} \; &= \; \tub[\tau]{\tau}{\tau}{\raisebox{-1.5pt}{$\scriptstyle 1$}} \, + \, a\cdot \tub[\tau]{\tau}{\tau}{\tau} \; ,
\end{aligned}
\end{equation}
where again $a=a_-$ for $\text{Fib}^+$ and $a = a_+$ for $\text{Fib}^-$. Associativity of the algebra multiplication as well as its compatibility with the $\ast$-structure can be checked to hold as a consequence of the identity $a^2 = a + 1$.

\textbf{Tube representations:} There is a total of four irreducible representations of $\text{Tube}(\text{Fib}^{\pm})$, which can be described as follows (here, $a = a_{\pm}$ for $\text{Fib}^{\mp}$):
\begin{itemize}
\item There is a one-dimensional representation $R_1$ acting on the twisted sector $\mathcal{H}_1 \cong \mathbb{C}$ via
\begin{equation}
R_1\Big( \tub[1]{1}{\tau}{} \Big) \; = \; - \, \frac{1}{a} \; .
\end{equation}
This is a $\ast$-representation for both $\text{Fib}^{\pm}$.
\item There are two one-dimensional representations $R_{\tau}^{\pm}$ acting on the twisted sector $\mathcal{H}_{\tau} \cong \mathbb{C}$ via
\begin{equation}
\begin{aligned}
R_{\tau}^{\pm}\Big( \tub[\tau]{\tau}{\tau}{\raisebox{-1.5pt}{$\scriptstyle 1$}} \Big) \; &= \; x_{\pm} \; , \\[3pt]
R_{\tau}^{\pm}\Big( \tub[\tau]{\tau}{\tau}{\tau} \Big) \; &= \; \frac{1}{a} \cdot \Big( 1- \frac{x_{\pm}}{a}\Big) \; ,
\end{aligned}
\end{equation}
where $x_{\pm}$ are the two solutions of $x^2+x+a^2=0$,
\begin{equation}
x_{\pm} \; = \; - \, \frac{1}{2} \, \pm \, i \cdot \sqrt{a + \frac{3}{4}}
\end{equation}
which are related by $x_- = (x_+)^{\ast}$ and $x_+ \cdot x_- = a^2$. These are $\ast$-representation for both $\text{Fib}^{\pm}$.
\item There is one two-dimensional representation $R_{1,\tau}$ acting on the twisted sectors $\mathcal{H}_1 \cong \mathcal{H}_{\tau} \cong \mathbb{C}$ via
\begin{equation}
\begin{aligned}
R_{1,\tau}\Big( \tub[1]{1}{\tau}{} \Big) \; &= \; \begin{pmatrix} \,a \; & \,0\, \\ \,0\; & \,0\, \end{pmatrix} \, , \\
R_{1,\tau}\Big( \tub[\tau]{\tau}{\tau}{\raisebox{-1.5pt}{$\scriptstyle 1$}} \Big) \; &= \; \begin{pmatrix} \,0 \; & \,0\, \\ \,0\; & \,-\,a\, \end{pmatrix}  \, ,\\
R_{1,\tau}\Big( \tub[\tau]{\tau}{\tau}{\raisebox{-1.5pt}{$\scriptstyle 1$}} \Big) \; &= \; \begin{pmatrix} \,0 \; & \,0\, \\ \,0\; & \,a^2\, \end{pmatrix}  \, ,\\
R_{1,\tau}\Big( \tub[1]{\tau}{\tau}{} \Big) \; &= \; \lambda \cdot \begin{pmatrix} \,0 \; & \,0\, \\ \, 1+ia \; & \,0\, \end{pmatrix}  \, ,\\
R_{1,\tau}\Big( \tub[\tau]{1}{\tau}{} \Big) \; &= \; \frac{1}{\lambda} \cdot \begin{pmatrix} \,0 \; & \,1-ia\, \\ \, 0 \; & \,0\, \end{pmatrix} \, .
\end{aligned}
\end{equation}
Here, $\lambda \in \mathbb{C}^{\times}$ is a gauge parameter that describes equivalent representations of $\text{Tube}(\text{Fib}^{\pm})$. One can check that these are $\ast$-representations if and only if
\begin{equation}
|\lambda|^2 \; = \; - \, \frac{1}{a} \; ,
\end{equation}
which admits a solution only when $a = a_-$ (in which case $-1/a_- = a_+ > 0$). Consequently, we see that the non-unitary Yang-Lee category $\text{Fib}^-$ has a tube representation which is not a $\ast$-representation.
\end{itemize}

\section{Three Dimensions}
\label{sec-three-dimensions}

In this section, we review the construction of the tube algebra $\text{Tube}(\mathcal{C})$ associated to a fusion 2-category $\mathcal{C}$ as introduced in \cite{Bartsch2023a}. We describe its canonical $\ast$-structure and classify irreducible $\ast$-representations in several examples related to higher group symmetries.

\subsection{Preliminaries}

In three dimensions, the finite (bosonic) generalised symmetries of a quantum field theory are described by a fusion 2-category $\mathcal{C}$ \cite{Douglas:2018qfz}, whose objects $A,B \in \mathcal{C}$ are topological surface defects, 1-morphisms $\varphi,\psi: A \to B$ are topological line interfaces and 2-morphisms $\Theta: \varphi \Rightarrow \psi$ are topological junctions:
\begin{equation}
\vspace{-5pt}
\begin{gathered}
\includegraphics[height=1.25cm]{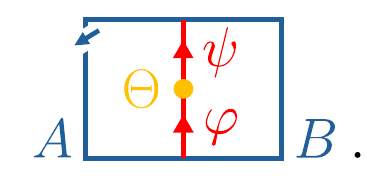}
\end{gathered}
\end{equation}
The monoidal structure $\otimes: \mathcal{C} \times \mathcal{C} \to \mathcal{C}$ captures the fusion of topological surfaces, their interfaces and junctions:
\begin{equation}
\vspace{-5pt}
\begin{gathered}
\includegraphics[height=1.65cm]{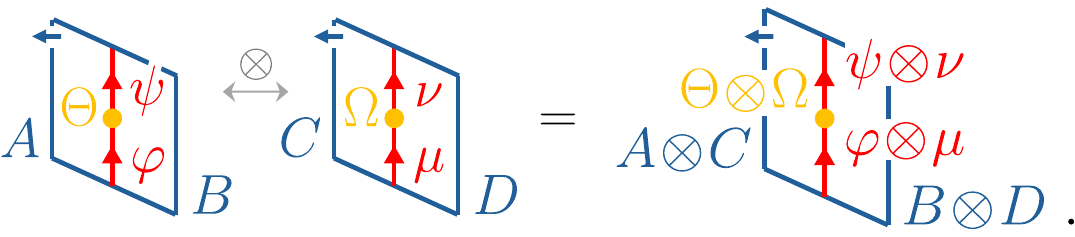}
\end{gathered}
\end{equation}
We further assume the following structures on $\mathcal{C}$:
\begin{enumerate}[label=\textbf{\arabic*}.]
\item \textbf{Dual structure:} A \textit{dual structure} on $\mathcal{C}$ allows us to bend both topological surfaces and lines around. Concretely, it consists of the following pieces of data:
\begin{itemize}[leftmargin=13pt]
\item A choice of \textit{(1-)dual object} $A^{\vee_1} \in \mathcal{C}$ for every object $A \in \mathcal{C}$ together with evaluation and coevaluation 1-morphisms
\begin{equation}
\vspace{-5pt}
\begin{gathered}
\;\;\qquad\includegraphics[height=1.55cm]{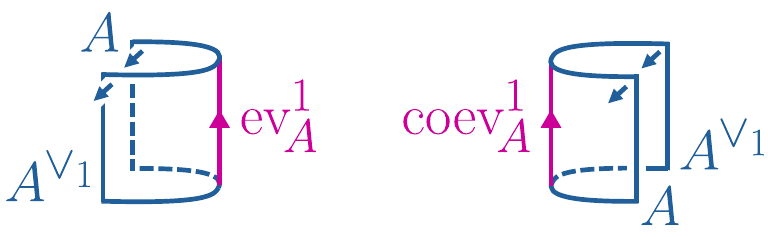}
\end{gathered}
\end{equation}
and cusp and casp 2-isomorphisms
\begin{equation}
\vspace{-5pt}
\begin{gathered}
\;\;\quad\includegraphics[height=1.9cm]{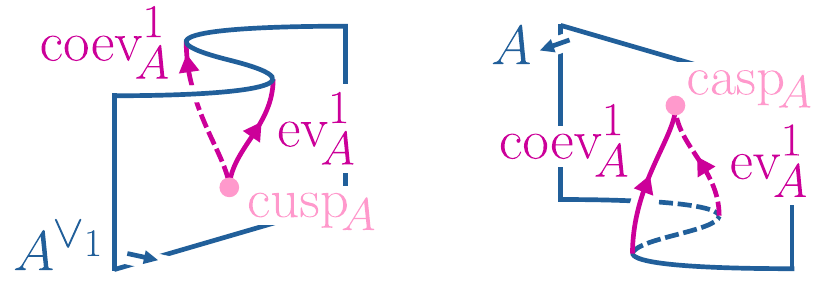}
\end{gathered}
\end{equation}
satisfying \textit{swallowtail relations} \cite{Douglas:2018qfz}. The assignment $A \mapsto A^{\vee_1}$ then extends to a 1op-monoidal 2-functor $\vee_1: \mathcal{C} \to \mathcal{C}^{\text{1op}}$. A \textit{1-pivotal structure} on $\mathcal{C}$ is a natural isomorphism $\xi: (\vee_1)^2 \Rightarrow \text{id}_{\mathcal{C}}$.
\item A choice of \textit{(2-)dual 1-morphism} $\varphi^{\vee_2}: B \to A$ for every 1-morphism $\varphi: A \to B$ in $\mathcal{C}$ together with evaluation and coevaluation 2-morphisms
\begin{equation}
\vspace{-5pt}
\begin{gathered}
\;\;\quad\includegraphics[height=1.4cm]{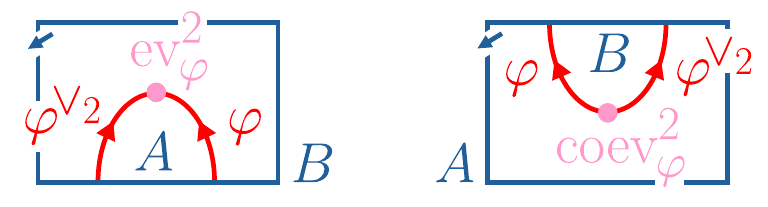}
\end{gathered}
\end{equation}
satisfying suitable \textit{snake} or \textit{zig-zag relations}. The assignment $\varphi \mapsto \varphi^{\vee_2}$ then extends to a monoidal 2-functor\footnote{We denote by $\mathcal{C}^{\text{2op}}$ the 2-category with the same objects and 1-morphisms as $\mathcal{C}$ and 2-morphisms given by $\text{2Hom}_{(\mathcal{C}^{\text{2op}})}(\varphi,\psi) = \text{2Hom}_{\mathcal{C}}(\psi,\varphi)$ for all 1-morphisms $\varphi$ and $\psi$.} $\vee_2: \mathcal{C} \to \mathcal{C}^{\text{2op}}$ acting as the identity on objects. A \textit{2-pivotal structure} on $\mathcal{C}$ is a natural isomorphism $\Xi: \vee_2 \circ \vee_2 \Rightarrow \text{id}_{\mathcal{C}}$ with trivial component 1-morphisms. Using this, the left and right traces of a 2-endomorphism $\Theta \in \text{2End}_{\mathcal{C}}(\varphi)$ of a 1-morphism $\varphi: A \to B$ are defined by\footnote{Note that $\text{tr}_L(\Theta) \in \text{2End}_{\mathcal{C}}(\text{id}_A)$ whereas $\text{tr}_R(\Theta) \in \text{2End}_{\mathcal{C}}(\text{id}_B)$.}
\begin{equation}
\vspace{-5pt}
\begin{gathered}
\;\;\quad\includegraphics[height=2.7cm]{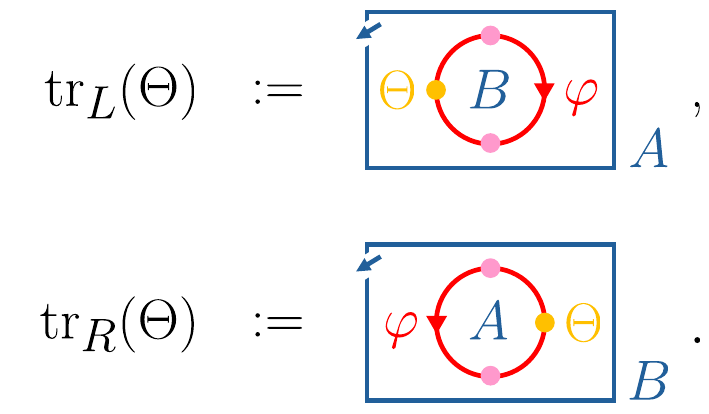}
\end{gathered}
\end{equation}
\end{itemize}
A dual structure on $\mathcal{C}$ is called \textit{spherical} if it is equipped with pivotal structures such that the associated front and back traces of 2-endomorphisms $\Theta \in \text{2End}_{\mathcal{C}}(\text{id}_A)$ agree \cite{Douglas:2018qfz}, i.e.
\begin{equation}
\vspace{-5pt}
\begin{gathered}
\;\;\quad\includegraphics[height=1.175cm]{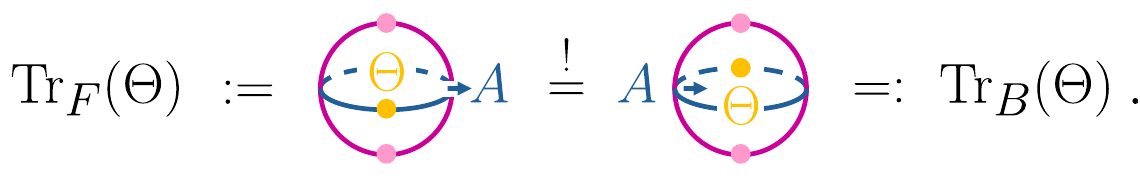}
\vspace{-6pt}
\end{gathered}
\vspace{8pt}
\end{equation}
In this case, we define the dimension of an object $A \in \mathcal{C}$ by $\text{dim}(A) := \text{Tr}_{F/B}(\text{Id}_{\hspace{1pt}\text{id}_A})$. Similarly, we set $\text{dim}(\varphi) := \text{Tr}_{F/B}(\text{tr}_{L/R}(\text{Id}_{\varphi}))$ for any morphism $\varphi$\footnote{Here, we use the fact that $\text{Tr}_F(\text{tr}_L(\Theta)) = \text{Tr}_F(\text{tr}_R(\Theta))$ for any 2-endomorphism $\Theta$ and similarly for the back trace \cite{Douglas:2018qfz}.}.

\item \textbf{Dagger structure:} A \textit{$\dagger$-structure} on $\mathcal{C}$ allows us to reflect topological junctions about a fixed hyperplane. Concretely, there is a 2-functor $\dagger: \mathcal{C} \to \mathcal{C}^{\text{2op}}$ that acts as the identity on objects and 1-morphisms and anti-linearly on 2-morphisms via
\begin{equation}
\vspace{-5pt}
\begin{gathered}
\includegraphics[height=1.25cm]{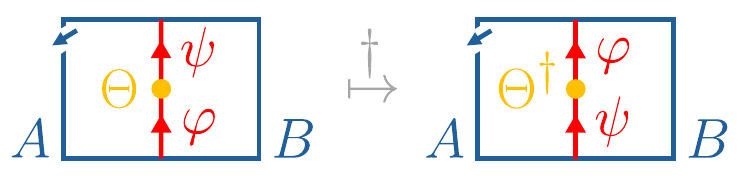}
\end{gathered}
\end{equation}
such that $\dagger^2 = \text{id}_{\mathcal{C}}$ and $\text{2End}_{\mathcal{C}}(\varphi)$ is a C*-algebra for all 1-morphisms $\varphi$ in $\mathcal{C}$. Furthermore, we assume all higher coherence data\footnote{We collectively refer to any natural coherence 2-isomorphisms as \textit{higher coherence data} in what follows. In the above case, the higher coherence data associated to $\mathcal{C}$ as a 2-category is given by the associator and unitors for the composition of 1-morphisms.} associated to $\mathcal{C}$ as a 2-category to be unitary. This turns $\mathcal{C}$ into a $\dagger$-2-category in the sense of \cite{Longo:1996hkk,Chen:2021ttc}. We assume the $\dagger$-structure to be compatible with the monoidal structure on $\mathcal{C}$ in the sense that $\otimes: \mathcal{C} \times \mathcal{C} \to \mathcal{C}$ is a $\dagger$-2-functor\footnote{A 2-functor $F: \mathcal{C} \to \mathcal{C}'$ between two $\dagger$-2-categories $\mathcal{C}$ and $\mathcal{C}'$ is called a \textit{$\dagger$-2-functor} if $\dagger' \circ F = F^{\text{2op}} \circ \dagger$ and all associated higher coherence data is unitary.} with unitary higher coherence data.
\end{enumerate}
As before, we assume the above structures to be compatible with one another in the sense that both $\vee_1: \mathcal{C} \to \mathcal{C}^{\text{1op}}$ and $\vee_2: \mathcal{C} \to \mathcal{C}^{\text{2op}}$ are $\dagger$-2-functors with unitary higher coherence data. In this case,
\begin{equation}
\vspace{-5pt}
\begin{gathered}
\includegraphics[height=1.95cm]{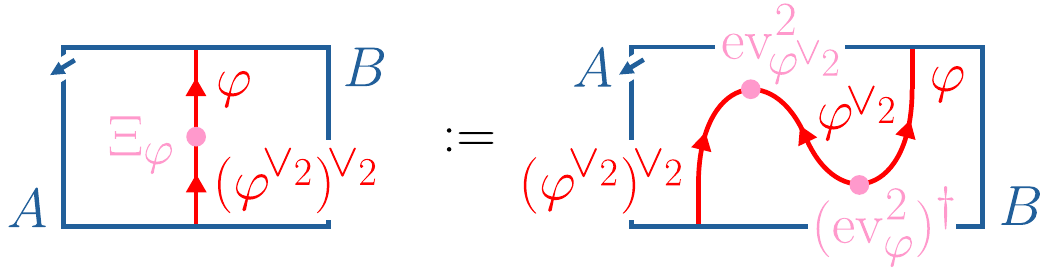}
\end{gathered}
\end{equation}
defines a canonical 2-pivotal structure on $\mathcal{C}$. Furthermore, we assume the 1-pivotal structure $\xi$ to have unitary higher coherence data. We call a dual structure of this type a \textit{unitary dual structure}. In what follows, we assume $\mathcal{C}$ to be a unitary fusion 2-category equipped with a spherical unitary dual structure.

\subsection{Tube Algebra}

Given the symmetry 2-category $\mathcal{C}$, we can associate to it its \textit{tube algebra} $\text{Tube}(\mathcal{C})$ \cite{Bartsch2023a}, which captures the linking action of topological surface defects on twisted sector local operators in three dimensions. As a vector space, the tube algebra decomposes into a direct sum
\begin{equation}
\text{Tube}(\mathcal{C}) \;\;\; = \bigoplus_{[\mu],[\nu] \hspace{1pt} \in \hspace{1pt} \pi_1(\mathcal{C})} \!\! \text{Tube}(\mathcal{C})_{\mu,\nu}
\end{equation}
where the summand associated to simple lines $\mu,\nu \in \Omega\mathcal{C}$ is given by the quotient space\footnote{For brevity, we often denote the identity 1-morphism of an object $A \in \mathcal{C}$ by the same letter $A$. Similarly for the identity 2-morphism of a 1-morphism $\varphi$ in $\mathcal{C}$.}
\begin{equation}
\text{Tube}(\mathcal{C})_{\mu,\nu} \; = \; \bigoplus_{A \hspace{1pt} \in \hspace{1pt} \mathcal{C}} \, \text{2Hom}_{\mathcal{C}}(A \otimes \mu \hspace{1pt}, \nu \otimes A) \; \Big/ \, \sim
\end{equation}
of local intersection 2-morphisms
\begin{equation}
\label{eq-tube-objects}
\vspace{-5pt}
\begin{gathered}
\includegraphics[height=1.7cm]{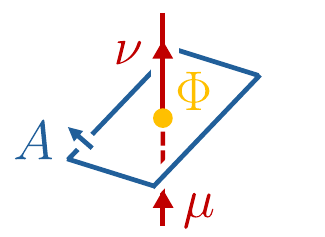}
\end{gathered}
\end{equation}
subjected to the equivalence relation generated by\footnote{Here, we implicitly make use of the 2-dual structure on $\mathcal{C}$.}
\begin{equation}
\vspace{-5pt}
\begin{gathered}
\includegraphics[height=2.1cm]{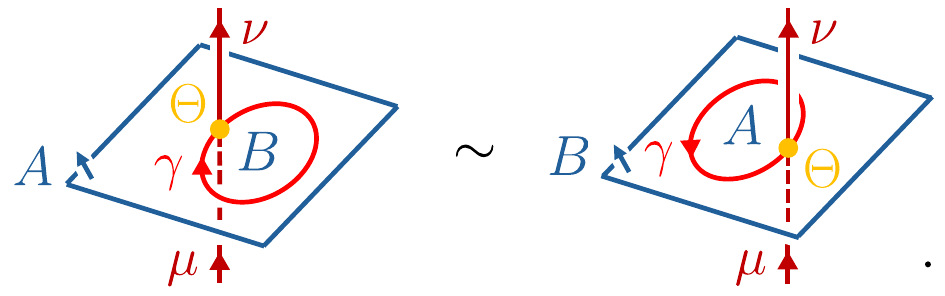}
\end{gathered}
\end{equation}
We denote the equivalence class of 2-morphisms (\ref{eq-tube-objects}) by
\begin{equation}
\htub[\mu]{\nu}{A}{\Phi} \; \in \; \text{Tube}(\mathcal{C})_{\mu,\nu} \; .
\end{equation}
As a vector space, $\text{Tube}(\mathcal{C})_{\mu,\nu}$ is isomorphic to
\begin{equation}
\text{Tube}(\mathcal{C})_{\mu,\nu} \; \cong \! \bigoplus_{[A] \hspace{1pt} \in \hspace{1pt} \pi_0(\mathcal{C})} \hspace{-6pt} \text{2Hom}_{\mathcal{C}}(A \otimes \mu \hspace{1pt}, \nu \otimes A) \; ,
\end{equation}
which shows that the tube algebra is finite-dimensional. The algebra multiplication is induced by
\begin{equation}
\vspace{-5pt}
\begin{gathered}
\includegraphics[height=2.25cm]{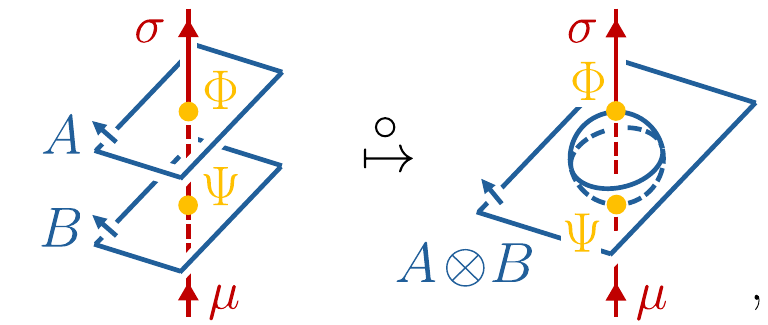}
\end{gathered}
\end{equation}
which we denote by
\begin{equation}
\htub[\hspace{-2pt}\nu']{\sigma}{A}{\Phi} \; \circ \; \htub[\mu]{\nu}{B}{\hspace{1pt}\Psi} \;\; = \;\; \delta_{\hspace{1pt}\nu,\nu'} \, \cdot \, \htuub[\mu]{\sigma}{\hspace{-2.5pt} A \hspace{0.5pt} \otimes B}{\hspace{1pt} \Phi \hspace{0.6pt}\circ\hspace{0.6pt} \Psi}
\end{equation}
on the generators of $\text{Tube}(\mathcal{C})$. The tube algebra further admits an antilinear involution $\ast: \text{Tube}(\mathcal{C}) \to \text{Tube}(\mathcal{C})$, which is induced by\footnote{Here, we implicitly make use of the 1-dual structure on $\mathcal{C}$.}
\begin{equation}
\label{eq-3d-tube-invol}
\vspace{-5pt}
\begin{gathered}
\includegraphics[height=1.85cm]{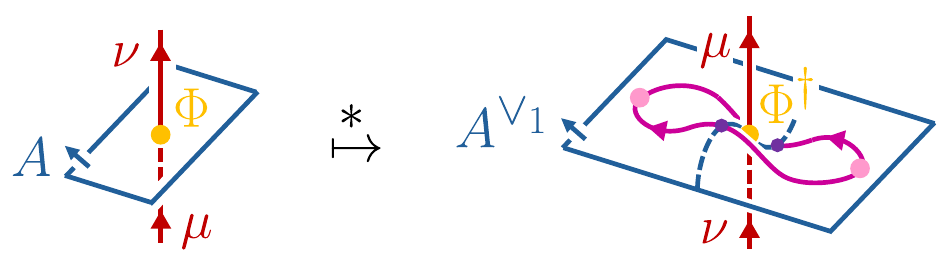}
\end{gathered}
\end{equation}
and which we denote by
\begin{equation}
\label{eq-tube-invol}
\htub[\mu]{\nu}{A}{\Phi}^{\ast} \; = \; \htub[\nu]{\mu}{\hspace{-2pt}A^{\hspace{-0.6pt}\vee_1}}{\hspace{-0.7pt}\raisebox{-1.1pt}{\hspace{1.2pt}$\scriptstyle \Phi^{\hspace{-0.5pt}\dagger}$}}
\end{equation}
on the generators of the tube algebra.

\subsection{Examples}

We conclude this section with examples of unitary fusion 2-category symmetries and $\ast$-representations of their associated tube algebras in the context of higher group symmetries in three dimensions.

\subsubsection{Group Symmetry}

Let us first consider the fusion 2-category\footnote{Here, we denote by $\text{2Hilb}$ the 2-category of 2-Hilbert spaces in the sense of \cite{BAEZ1997125}, viewed as a $\dagger$-2-category in the sense of \cite{Longo:1996hkk,Chen:2021ttc}.} $\mathcal{C} = \text{2Hilb}_G^{\pi}$ corresponding to a finite group symmetry $G$ with 't Hooft anomaly $[\pi] \in H^4(G,U(1))$. In this case, the associated tube algebra is simply the group algebra $\mathbb{C}[G]$ of $G$ with $\ast$-structure given by inversion \cite{Bartsch2023a}. In particular, $\ast$-representations of the tube algebra are unitary representations of $G$. We now aim to rederive this result using higher $S$-matrices.

\textbf{Drinfeld centre:} As shown in \cite{Kong2020}, the Drinfeld centre of $\mathcal{C}$ (as a linear 2-category) decomposes as
\begin{equation}
\label{eq-drinfeld-centre-anomalous-group}
\mathcal{Z}(\text{2Hilb}_G^{\pi}) \;\;  = \bigoplus_{[x] \, \in \, \text{Cl}(G)} \text{2Rep}^{\tau_x(\pi)}(G_x) \; ,
\end{equation}
so that the simple objects can be labelled by pairs $(x,\sigma)$ consisting of a representative $x \in G$ of a conjugacy class $[x] \in \text{Cl}(G)$ together with an irreducible 2-representation \cite{GANTER20082268,ELGUETA200753,OSORNO2010369} $\sigma$ of the centraliser $G_x$ of $x$ with projective 3-cocycle $\tau_x(\pi) \in Z^3(G_x,U(1))$. Here, we defined the \textit{transgression} of the 't Hooft anomaly $\pi$ by 
\begin{equation}
\tau_x(\pi)(g,h,k) \; := \; \frac{\pi(g,h,k,x) \cdot \pi(g,{}^{hk}x,h,k)}{\pi(g,h,{}^kx,k) \cdot \pi({}^{ghk}x,g,h,k)} \; .
\end{equation}
The monoidal unit of $\mathcal{Z}(\mathcal{C})$ is given by $\mathbf{1} = (e,\one)$, where $e \in G$ is the identity element and $\one$ is the trivial 2-repre-sentation of $G$. In order to compute the connected components, we use the fact that 
\begin{equation}
\text{2Rep}^{\alpha}(H) \; = \; \text{Mod}(\text{Hilb}_H^{\hspace{0.7pt}\alpha})
\end{equation}
for any finite group $H$ and 3-cocycle $\alpha \in Z^3(H,U(1))$, which, since $\text{Mod}(\mathcal{B})$ is connected for any $\mathcal{B}$ \cite{Douglas:2018qfz,Johnson-Freyd:2021chu}, shows that the connected components of $\mathcal{Z}(\mathcal{C})$ are in 1:1-correspondence with the conjugacy classes of $G$, i.e.
\begin{equation}
\pi_0\big( \mathcal{Z}(\text{2Hilb}_G^{\pi}) \big) \; = \; \text{Cl}(G) \; .
\end{equation}
Moreover, using that gauging a $G$-symmetry in three dimensions leads to topological Wilson lines \cite{Bartsch:2022mpm,Bhardwaj:2022lsg,Bartsch:2022ytj,Bhardwaj:2022kot,Bhardwaj:2022maz},
\begin{equation}
\text{End}_{\hspace{1pt}\text{2Rep}(G)}(\one) \; =  \; \text{Rep}(G) \; ,
\end{equation} 
we see that the fundamental hypergroup of $\mathcal{Z}(\mathcal{C})$ is
\begin{equation}
\pi_1\big(\mathcal{Z}(\text{2Hilb}_G^{\pi})\big) \; = \; \text{Irr}(\text{Rep}(G)) \; .
\end{equation}
In particular, since the number of conjugacy classes equals the number of isomorphism classes of irreducible representations, we see that $|\pi_0(\mathcal{Z}(\mathcal{C}))| = |\pi_1(\mathcal{Z}(\mathcal{C}))|$.

\textbf{S-matrix:} The $S$-matrix of $\mathcal{Z}(\mathcal{C})$ is the square matrix
\begin{equation}
S: \; \text{Cl}(G) \hspace{1pt}\times\hspace{1pt} \text{Irr}(\text{Rep}(G)) \; \to \; \mathbb{C}
\end{equation}
that computes the character table of $G$ \cite{Reutter2022}, i.e.
\begin{equation}
S_{\hspace{1pt}[x],[\rho]} \; = \; \frac{\text{Tr}(\rho(x))}{\text{dim}(\rho)} \; .
\end{equation}
In order to simplify notation, we set
\begin{equation}
n \; := \; |\text{Cl}(G)| \; = \; |\text{Irr}(\text{Rep})(G)|
\end{equation}
and fix for each $i \in \lbrace 1,...,n \rbrace$ a representative $x_i \in G$ of the corresponding conjugacy class $[x_i] \in \text{Cl}(G)$ as well as a representative $\rho_i \in \text{Rep}(G)$ of the corresponding isomorphism class $[\rho_i] \in \text{Irr}(\text{Rep}(G))$ of irreducible representations of $G$. Furthermore, we denote by $G_i := G_{x_i}$ the centraliser of $x_i$ and by $\chi_i := \text{Tr}(\rho_i(.))$ the character associated to the irreducible representation $\rho_i$ (whose dimension we denote by $d_i := \text{dim}(\rho_i)$). Using this, the $S$-matrix is the $(n \times n)$-matrix with entries
\begin{equation}
S_{ij} \; := \; S_{\hspace{1pt}[x_i],[\rho_j]} \; \equiv \; \frac{\chi_j(x_i)}{d_j} \; .
\end{equation} 
As a consequence of the properties of characters such as the character orthogonality relations
\begin{align}
\sum_{k\hspace{1pt}=\hspace{1pt}1}^n \, \frac{1}{|G_k|} \cdot \chi^{\ast}_i(x_k) \cdot \chi_j(x_k) \; &= \; \delta_{ij} \; , \\
\frac{1}{|G_i|} \cdot \sum_{k\hspace{1pt}=\hspace{1pt}1}^n \, \chi^{\ast}_k(x_i) \cdot \chi_k(x_j) \; &= \; \delta_{ij} \; , 
\end{align}
the $S$-matrix has the following properties:
\begin{enumerate}[label=\protect\circled{\arabic*}]
\item It is invertible with inverse given by
\begin{equation}
(S^{-1})_{ij} \; = \; \frac{d_i}{|G_j|} \cdot \chi^{\ast}_i(x_j) \; .
\end{equation}
\item It satisfies $S_{\hspace{1pt}i^{\vee}\!,\hspace{1pt}j} = (S_{ij})^{\ast}$, where $i^{\vee} \in \lbrace 1,...,n \rbrace$ is such hat $[x_{(i^{\vee})}] = [(x_i)^{-1}] \in \text{Cl}(G)$.
\item It satisfies the Verlinde formula
\begin{equation}
S_{i\ell} \cdot S_{j\ell} \; = \; \sum_{k\hspace{1pt}=\hspace{1pt}1}^n N_{ij}^k \cdot S_{k\ell} \; ,
\end{equation}
where the hypergroup coefficients $N_{ij}^k \in \mathbb{C}$ are 
\begin{equation}
\begin{gathered}
\qquad N_{ij}^k \; = \\[-2pt]
\qquad \frac{1}{|G_k|} \cdot \sum_{p\hspace{1pt}=\hspace{1pt}1}^n \, \frac{1}{d_p} \cdot \chi_p(g_i) \cdot \chi_p(g_j) \cdot \chi^{\ast}_p(g_k) \; .
\end{gathered}
\end{equation}
\end{enumerate}

\textbf{Tube idempotents:} As described in section \ref{ssec-overview}, the $S$-matrix allows us to establish an equivalence between the category of $\ast$-representations of the tube algebra and
\begin{equation}
\Omega(\mathcal{Z}(\text{2Hilb}_G^{\pi})) \; = \; \text{Rep}(G) 
\end{equation}
via minimal central idempotents. Concretely, using (\ref{eq-min-cen-idems}), the minimal central idempotent associated to an irreducible representation $\rho \in \text{Rep}(G)$ is computed to be
\begin{equation}
e_{\rho} \,\; = \;\, \frac{\text{dim}(\rho)}{|G|} \cdot \sum_{g \hspace{1pt} \in \hspace{1pt} G} \chi_{\rho}^*(g) \cdot g \; ,
\end{equation}
which reproduces the well-known formula for minimal central idempotents in the group algebra $\mathbb{C}[G]$ \cite{Janusz1966}.

\subsubsection{2-Group Symmetry}

As a second example, let us consider $\mathcal{C} = \text{2Hilb}_{\mathcal{G}}^{\lambda}$ corresponding to a finite anomalous 2-group symmetry \cite{Baez2004,Benini:2018reh} $\mathcal{G} = A[1] \rtimes_{\alpha} G$ that is specified by the following data:
\begin{enumerate}
\item a finite 0-form symmetry group $G$,
\item a finite abelian 1-form symmetry group $A$,
\item a group action\footnote{We will henceforth denote the action of a group element $g \in G$ on $a \in A$ by $g \triangleright a =: {}^{g\hspace{-1pt}}a$. Furthermore, we set $a^g := (g^{-1}) \triangleright a$.} $\triangleright: G \to \text{Aut}(A)$,
\item a Postnikov class $[\alpha] \in H^3_{\hspace{1pt}\triangleright}(G,A)$,
\item a mixed 't Hooft anomaly $[\lambda] \in H^2_{\hspace{1pt}\triangleright}(G,A^{\vee})$.
\end{enumerate}

\vspace{4pt}
\textbf{Tube algebra:} As demonstrated in \cite{Bartsch2023a}, the tube algebra associated to $\text{2Hilb}_{\mathcal{G}}^{\lambda}$ is the twisted groupoid algebra $\mathbb{C}^{\varepsilon(\lambda)}[A/\!/G]$ with generators $\vphantom{\rule{0pt}{10pt}}\raisebox{-1pt}{$\htubb[a]{\hspace{-2.7pt}{}^{\mathrlap{\raisebox{-2.5pt}{\hspace{-5pt}\crule[white]{8pt}{5pt}}}g}\hspace{-0.95pt}a}{g}{}$}$ ($g\in G,\, a \in A$) that multiply according to
\begin{equation}
\label{eq-2-grp-alg-mult}
\htubb[a]{\hspace{-2.7pt}{}^{\mathrlap{\raisebox{-2.5pt}{\hspace{-5pt}\crule[white]{8pt}{8pt}}}g}\hspace{-0.95pt}a}{g}{} \hspace{1pt}\circ\hspace{1pt} \htubb[b]{\hspace{-2.7pt}{}^{\mathrlap{\raisebox{-2.5pt}{\hspace{-5pt}\crule[white]{8pt}{8pt}}}h}\hspace{-0.6pt}b}{\hspace{-0.4pt}h}{} \;\, = \,\; \delta_{a,{}^h\hspace{-0.7pt}b} \hspace{1pt}\cdot\hspace{1pt} \varepsilon_b(g,h) \hspace{1pt}\cdot\hspace{1pt} \htubb[b]{\hspace{-6pt}{}^{\mathrlap{\raisebox{-2.5pt}{\crule[white]{6pt}{6pt}}} g\hspace{-0.5pt}h}\hspace{-0.8pt}b}{\hspace{-2pt}g\hspace{-0.4pt}h}{} \; ,
\end{equation}
where we defined the multiplicative phase
\begin{equation}
\varepsilon_a(\lambda)(g,h) \; = \; \braket{\lambda(g,h),{}^{gh}a} \; .
\end{equation}
As a result of the cocycle condition for $\lambda$, it obeys
\begin{equation}
\label{eq-3d-transgression-cocycle-cond}
(d\varepsilon)_a(g,h,k) \;\, := \;\, \frac{\varepsilon_a(h,k) \cdot \varepsilon_a(g,hk)}{\varepsilon_a(gh,k) \cdot \varepsilon_{({}^ka)}(g,h)} \,\; = \;\, 1 \, ,
\end{equation}
which ensures that the algebra multiplication in (\ref{eq-2-grp-alg-mult}) is associative. Using (\ref{eq-3d-tube-invol}), the $\ast$-structure can be computed to be
\begin{equation}
\label{eq-2-group-invol}
\htubb[a]{\hspace{-2.7pt}{}^{\mathrlap{\raisebox{-2.5pt}{\hspace{-5pt}\crule[white]{8pt}{8pt}}}g}\hspace{-0.95pt}a}{g}{}^{\ast} \,\; = \,\; \mu_a(g) \hspace{1pt} \cdot \hspace{1pt} \htubb[\hspace{-2.3pt}{}^g\hspace{-0.95pt}a]{\hspace{-0.5pt}a}{\hspace{-0.1pt}g^{-1}}{} \; ,
\end{equation}
where we defined the multiplicative phase
\begin{equation}
\mu_a(g) \; := \; \varepsilon^{\ast}_a(g^{-1},g) \; .
\end{equation}
As a consequence of (\ref{eq-3d-transgression-cocycle-cond}), it satisfies
\begin{equation}
\mu_a(g^{-1}) \; = \; \mu_{(a^g)}(g) \quad\;\; \text{and} \quad\;\; d\mu \; = \; \widehat{\varepsilon} \hspace{1pt}/ \hspace{1pt} \varepsilon \; ,
\end{equation}
where we defined the 2-cocycle
\begin{equation}
\widehat{\varepsilon}_a(g,h) \; := \; \varepsilon^{\ast}_{({}^{gh}a)}(h^{-1},g^{-1}) \; .
\end{equation}
This ensures that the $\ast$-structure (\ref{eq-2-group-invol}) is involutory and compatible with the algebra multiplication. As in the case of an ordinary group, we now aim to construct $\ast$-re-presentations of this algebra using higher $S$-matrices.

\textbf{Drinfeld centre:} In order to determine the Drinfeld centre of  $\mathcal{C} = \text{2Hilb}_{\mathcal{G}}^{\lambda}$, we use the fact that we can gauge the 1-form symmetry $A$ to obtain a pure 0-form symmetry $\widehat{G}$ given by the group extension \cite{Bartsch:2022mpm,Bhardwaj:2022lsg,Bartsch:2022ytj,Bhardwaj:2022kot,Bhardwaj:2022maz,Decoppet2023}
\begin{equation}
\widehat{G} \; := \; A^{\vee} \rtimes_{\lambda} G \; ,
\end{equation}
which as a set is $A^{\vee} \times G$ with multiplication defined by
\begin{equation}
(\mu,g) \cdot (\nu,h) \; := \; \big( \mu \cdot {}^g\nu \cdot \lambda(g,h)\hspace{0.7pt} , \, g \hspace{-1pt}\cdot\hspace{-1pt} h \big) \; .
\end{equation} 
This symmetry then has an 't Hooft anomaly parameterised by the 4-cocycle $\braket{\hspace{1.2pt}.\hspace{1.2pt},\alpha} \in Z^4(\widehat{G},\mathbb{C}^{\times})$ defined by
\begin{equation}
\begin{aligned}
&\braket{\hspace{1.2pt}.\hspace{1.2pt},\alpha}\hspace{-1pt}\big[ \hspace{1pt} (\mu,g), \hspace{1pt} (\nu,h), \hspace{1pt} (\varphi,k), \hspace{1pt} (\psi,\ell) \hspace{1pt} \big] \\[3pt] 
&\hspace{80pt}:= \; \big\langle \hspace{0.5pt} {}^{ghk}\psi, \hspace{1pt} \alpha(g,h,k) \hspace{0.5pt}\big\rangle \; .
\end{aligned}
\end{equation}
Using the gauge invariance of the Drinfeld centre,
\begin{equation}
\mathcal{Z}\big(\text{2Hilb}_{\mathcal{G}}^{\lambda}\big) \; = \; \mathcal{Z}\!\left(\text{2Hilb}_{\widehat{G}}^{\braket{\hspace{1.2pt}.\hspace{1.2pt},\alpha}}\right) \, ,
\end{equation}
we can then trace back the centre of the 2-group symmetry $\mathcal{G}$ to the computation of the centre of an ordinary group symmetry as in (\ref{eq-drinfeld-centre-anomalous-group}). Concretely, the homotopy components of $\mathcal{Z}(\text{2Hilb}_{\mathcal{G}}^{\lambda})$ can be described as follows:
\begin{itemize}
\item The connected components of $\mathcal{Z}(\text{2Hilb}_{\mathcal{G}}^{\lambda})$ are in one-to-one correspondence with the conjugacy classes of $\widehat{G} = A^{\vee} \!\rtimes_{\lambda} \!G$. To describe the latter, we note that 
\begin{equation}
\label{eq-conjugacy-class-extension}
\qquad {}^{(\mu,g)}(\chi,x) \; = \; 
\left( \, {}^g\chi \cdot \big(\mu \hspace{1pt}/\hspace{1pt} {}^{\raisebox{-0.62pt}{$\scriptstyle ({}^{\raisebox{-1pt}{$\scriptscriptstyle g$}\hspace{-0.7pt}}x)\hspace{-0.7pt}$}}\mu\big) \cdot \tau_x(g) \, , \, {}^gx \, \right)
\end{equation}
for any $(\mu,g),(\chi,x) \in \widehat{G}$, where we defined the \textit{transgression} of the 't Hooft anomaly $\lambda$ by
\begin{equation}
\tau_x(\lambda)(g) \; := \; \frac{\lambda(g,x)}{\lambda({}^gx,g)} \; .
\end{equation}
As a consequence of the twisted 2-cocycle condition obeyed by $\lambda$, it satisfies
\begin{equation}
\begin{gathered}
\hspace{25pt}(d\tau)_x(g,h) \; := \\[3pt] 
\quad \hspace{15pt}\frac{{}^g[\hspace{0.5pt}\tau_x(h)\hspace{0.5pt}] \hspace{1pt} \cdot \hspace{1pt} \tau_{({}^hx)}(g)}{\tau_x(gh)} \; = \; \frac{\lambda(g,h)}{{}^{({}^{gh}x)}\lambda(g,h)} \; .
\end{gathered}
\end{equation}
Upon defining the subgroup $A_x := \lbrace a \in A \, | \, {}^xa = a \rbrace$ and using the canonical identification
\begin{equation}
\frac{A^{\vee}}{\lbrace \hspace{1pt} \mu\hspace{1pt}/\hspace{1pt}{}^{x\hspace{-1pt}}\mu  \; | \; \mu \!\in\! A^{\vee} \rbrace} \; \cong \; (A_x)^{\vee} \; ,
\end{equation}
this shows that we can label the connected components of $\mathcal{Z}(\text{2Hilb}_{\mathcal{G}}^{\lambda})$ by equivalence classes of pairs $(x,\chi)$ consisting of a group element $x \in G$ and a character $\chi \in (A_x)^{\vee}$, where two such pairs $(x,\chi)$ and $(x',\chi')$ are considered equivalent if there exists a $g \in G$ such that 
\begin{equation}
x' \; = \; {}^gx \; , \qquad \chi' \; = \; {}^g\chi \cdot \tau_x(g) \; .
\end{equation}
We will denote the equivalence class of $(x,\chi)$ by $[x,\chi]$ in what follows.

\item The fundamental hypergroup of $\mathcal{Z}(\text{2Hilb}_{\mathcal{G}}^{\lambda})$ is given by the set of isomorphism classes of irreducible representations of $\widehat{G} = A^{\vee}\! \rtimes_{\lambda}\! G$. By induction, these can be labelled by pairs $(a,\rho)$ consisting of
\begin{enumerate}
\item a group element $a \in A$, viewed as a character on the Pontryagin dual group $A^{\vee}$,
\item an irreducible representation $\rho$ of the stabiliser $G_a := \lbrace g \in G \, | \, {}^ga = a \rbrace$ of $a$ with projective 2-co-cycle $\braket{\lambda,a} \in Z^2(G_a,U(1))$.
\end{enumerate}
The corresponding irreducible representation $\widehat{\rho}$ of $\widehat{G}$ is then given by 
\begin{equation}
\widehat{\rho} \; = \; \text{Ind}_{\hspace{1pt}\widehat{G}_a}^{\hspace{1pt}\widehat{G}}\!\hspace{-1pt}(a \otimes \rho) \; ,
\end{equation}
where we defined $\widehat{G}_a := A^{\vee}\! \rtimes_{\lambda} \!G_a$. Two such pairs $(a,\rho)$ and $(a',\rho')$ are considered equivalent if $\widehat{\rho}$ and $\widehat{\rho}^{\hspace{2pt}\prime}$ are equivalent as representations of $\widehat{G}$. More concretely, $(a,\rho)$ and $(a',\rho')$ are equivalent if there exists a $g \in G$ such that 
\begin{equation}
\label{eq-3d-fund-comp-equiv}
\qquad a' \; = \; {}^{g\hspace{-0.5pt}}a \; , \qquad \rho' \,\; \cong \,\; \braket{\sigma_g\hspace{0.5pt},\hspace{-0.5pt}{}^{g\hspace{-0.5pt}}a} \hspace{1pt} \otimes \hspace{1pt} {}^{g\!}\rho \; ,
\end{equation}
where we defined the multiplicative factor
\begin{equation}
\sigma_g(\lambda)(h) \; := \; \frac{\lambda(h,g)}{\lambda(g,h^{\hspace{0.5pt}g})} \; .
\end{equation}
As a result of the 2-cocycle condition for $\lambda$, it obeys
\begin{equation}
\begin{gathered}
\qquad\;\; (d\sigma)_g(h,k) \; := \\[3pt] 
\quad \hspace{15pt}\frac{{}^h[\sigma_g(k)] \hspace{1pt} \cdot \hspace{1pt} \sigma_g(h)}{\sigma_g(hk)} \; = \; \frac{\lambda(h,k)}{{}^{\raisebox{1pt}{$\scriptstyle g$}\hspace{-1pt}}\lambda(h^{\hspace{0.7pt}g},k^{\hspace{0.7pt}g})} \; ,
\end{gathered}
\end{equation}
which ensures that (\ref{eq-3d-fund-comp-equiv}) defines an equivalence relation. We will denote the equivalence class of $(a,\rho)$ by $[a,\rho]$ in what follows.
\end{itemize}

\textbf{S-matrix:} Using the above, the $S$-matrix associated to $\mathcal{Z}(\text{2Hilb}_{\mathcal{G}}^{\lambda})$ is the square matrix that is indexed by equivalence classes $[x,\chi]$ and $[a,\rho]$ with entries
\begin{equation}
\label{eq-2-group-s-matrix}
S_{\hspace{1pt}[x,\chi],[a,\rho]} \; = \; \frac{\text{Tr}_{\hspace{0.3pt}\widehat{\rho}\hspace{0.3pt}}(\hspace{0.5pt}\widehat{x}\hspace{0.5pt})}{\text{dim}(\widehat{\rho}\hspace{0.7pt})} \; ,
\end{equation}
where $\widehat{\rho} = \text{Ind}_{\hspace{1pt}\widehat{G}_a}^{\hspace{1pt}\widehat{G}}\!\hspace{-1pt}(a \otimes \rho) \in \text{Rep}(\widehat{G})$ and $\widehat{g} = (\chi,x) \in \widehat{G}$. Using the character formula for induced representations,
\begin{equation}
\text{Tr}_{\hspace{0.3pt}\widehat{\rho}\hspace{0.3pt}}(\hspace{0.5pt}\widehat{x}\hspace{0.5pt}) \;\; = \;\; \frac{1}{\raisebox{-3pt}{$|\widehat{G}_a|$}} \, \cdot \!\! \sum_{\substack{\; \widehat{g} \hspace{1pt} \in \hspace{1pt} \widehat{G} \hspace{1pt}: \\[1pt] {}^{\widehat{g}}\widehat{x} \hspace{1pt} \in \hspace{1pt} \widehat{G}_a}} \text{Tr}_{\hspace{0.7pt} a \otimes \rho}\big(\hspace{0.5pt}{}^{\widehat{g}\hspace{0.5pt}}\widehat{x}\hspace{0.5pt}\big) \; ,
\end{equation}
as well as the character orthogonality relation
\begin{equation}
\frac{1}{|A^{\vee}|} \hspace{1pt} \cdot \!\! \sum_{\mu \hspace{1pt} \in \hspace{1pt} A^{\vee}} \mu^{\ast}(b) \cdot \mu(c) \,\; = \;\, \delta_{\hspace{0.5pt}b,c} \; ,
\end{equation}
we can then compute that the $S$-matrix is given by
\begin{equation}
\refstepcounter{equation}
\begin{gathered}
\;\;\; S_{\hspace{1pt}[x,\chi],[a,\rho]} \,\; = \\ 
\frac{1}{\text{dim}(\rho) \hspace{-1pt}\cdot \hspace{-1pt} |G|} \; \cdot \!\! \sum_{\substack{\; g \hspace{1pt} \in \hspace{1pt} G \hspace{1pt}: \\[1pt] {}^{g\hspace{-1pt}}x \hspace{1pt} \in \hspace{1pt} G_a}} \!\braket{\tau_x(g),a} \hspace{1pt}\cdot \hspace{1pt} \text{Tr}_{\rho}({}^{g\hspace{-0.7pt}}x) \hspace{1pt} \cdot \hspace{1pt} \chi(a^g) \; .
\end{gathered}
\tag*{\raisebox{-2.2em}{(\theequation)}}
\end{equation}
In particular, it satisfies properties $\circled{1}-\circled{3}$, as is clear from its expression as a character table (\ref{eq-2-group-s-matrix}).

\textbf{Tube representations:} Using the above expression for the $S$-matrix and (\ref{eq-min-cen-idems}), we can compute the minimal central idempotents in $\mathbb{C}^{\varepsilon(\lambda)}[A/\!/G]$ to be
\begin{equation}
\begin{gathered}
\quad e_{(a,\rho)} \;\; = \\ 
\frac{\text{dim}(\rho)}{|G_a|^2} \; \cdot \!\! \sum_{\substack{\; x,\hspace{0.5pt}g \hspace{1pt} \in \hspace{1pt} G \hspace{1pt}: \\[1pt] {}^{g\hspace{-1pt}}x \hspace{1pt} \in \hspace{1pt} G_a}} \!\! \braket{\tau_x(g),a}^{\ast} \cdot \text{Tr}_{\rho}^{\ast}({}^gx) \hspace{1pt}\cdot\hspace{1pt} \htubb[\hspace{-2.3pt}{}^g\hspace{-1.3pt}a]{\hspace{-2.7pt}{}^{\mathrlap{\raisebox{-2.5pt}{\hspace{-5pt}\crule[white]{8pt}{8pt}}}g}\hspace{-1.3pt}a}{\raisebox{-1pt}{$\scriptstyle x$}}{} \; ,
\end{gathered}
\end{equation}
where, as before, the label $(a,\rho)$ consists of 
\begin{enumerate}
\item a group element $a \in A$,
\item an irreducible representation $\rho$ of $G_a$ with projective 2-cocycle $\varepsilon_a(\lambda) \in Z^2(G_a,U(1))$.
\end{enumerate}
In particular, since there is a one-to-one correspondence between minimal central idempotents and irreducible representations of the tube algebra, we see that the latter are labelled by the same data $(a,\rho)$. Concretely, the irreducible representation $R_{\hspace{1pt}(a,\rho)}$ corresponding to $(a,\rho)$ can be constructed via induction: To this end, we fix for each $b \in [a]$ in the $G$-orbit $[a] \subset A$ a representative $r_b \in G$ such that ${}^{(r_b)}b = a$ (with $r_a = e$). Using these, we can define
\begin{equation}
g_b \; := \; r_{({}^gb)} \cdot g \cdot r_b^{-1} \; \in \; G_a
\end{equation}
for all $g \in G$ and $b \in [a]$. If we denote by $\mathcal{V}$ the Hilbert space underlying the (projective) representation $\rho$ of $G_a$, then $R_{(a,\rho)}$ acts on the Hilbert space $\mathcal{H}$ with $A$-grading
\begin{equation}
\mathcal{H}_b \; = \; \begin{cases} \mathcal{V} &\text{if} \; b \in [a] \\ 0 &\text{otherwise} \end{cases}
\end{equation}
via the induced tube action
\begin{equation}
R_{(a,\rho)}\Big( \htubb[b]{\hspace{-2.7pt}{}^{\mathrlap{\raisebox{-2.5pt}{\hspace{-2pt}\crule[white]{5pt}{8pt}}}g}\hspace{-0.7pt}b}{g}{} \Big) \; := \; \kappa_b(g) \cdot \rho( g_b ) \; ,
\end{equation}
where we defined the multiplicative phase
\begin{equation}
\kappa_b(g) \; := \; \frac{\varepsilon_b(r_{({}^gb)},g)}{\varepsilon_b(g_b,r_b)} \; .
\end{equation}
As a consequence of (\ref{eq-3d-transgression-cocycle-cond}), it satisfies
\begin{equation}
\frac{\kappa_b(h) \cdot \kappa_{({}^hb)}(g)}{\kappa_b(gh)} \; = \; \frac{\varepsilon_b(g,h)}{\varepsilon_a(g_{({}^hb)}, h_b)} \; ,
\end{equation}
which ensures that $R_{(a,\rho)}$ respects the algebra multiplication. As in the two-dimensional case, it is straightforward to check that $R_{(a,\rho)}$ is a $\ast$-representation if and only if $\rho$ is a unitary projective representation of $G_a$. Since every finite-dimensional representation of a finite group is equivalent to a unitary one, this shows that all irreducible representations of the tube algebra are $\ast$-representations as expected.

\section{Discussion}

In this note, we generalised the notion of unitary actions of global symmetry groups on local operators to the action of higher fusion category symmetries $\mathcal{C}$ on twisted sector operators in arbitrary spacetime dimension. We motivated that the latter transform $\ast$-representations of the tube algebra $\text{Tube}(\mathcal{C})$ associated to $\mathcal{C}$, which is expected to be a C*-algebra provided that the symmetry category $\mathcal{C}$ is unitary in an appropriate sense. In particular, $\mathcal{C}$ needs to be equipped with a $\dagger$-structure that implements reflections of top level morphisms (local defects) in $\mathcal{C}$. We then classified all irreducible $\ast$-representations of $\text{Tube}(\mathcal{C})$ using the Symmetry TFT and its associated higher S-matrices.

A natural question to ask is how the above generalises to the action of higher fusion category symmetries on extended observables. In the case of invertible symmetries, this was discussed in \cite{Bartsch2024}, where line operators in $D > 2$ were shown to transform in unitary 2-representations of a global 2-group symmetry. This required the usage of higher $\dagger$-structures that implement reflections of topological defects beyond the top level of morphisms. In the general case, we hence expect the symmetry category $\mathcal{C}$ to be equipped with a higher $\dagger$-structure that is compatible with the higher duality and coherence data associated to $\mathcal{C}$ \cite{Ferrer:2024vpn}. As a result, the higher tube algebras of $\mathcal{C}$ \cite{Bartsch2023a} are expected to inherit the structure of higher C*-algebras, whose higher $\ast$-representations capture the action of $\mathcal{C}$ on extended observables. We leave the exploration of these structures to future work.

More generally, one can consider unitary actions of the higher fusion category symmetry $\mathcal{C}$ on boundary operators sitting on topological boundary conditions for $\mathcal{C}$. In two dimensions, this was studied in \cite{Cordova2024,Choi2024} (see also \cite{Bhardwaj2024} for a Symmetry TFT perspective and \cite{Huang2023,Copetti2024,Copetti2025} for earlier work on related aspects), where boundary changing local operators were shown to transform in $\ast$-representations of a generalised tube (or strip) algebra associated to $\mathcal{C}$. Generalising these constructions to higher dimensions will be the subject of future of work.


\onecolumngrid
\vspace{18pt}

\begin{center}
\begin{minipage}{0.7\textwidth}
\centering
\textbf{ACKNOWLEDGEMENTS}

\vspace{5pt}
The author thanks Mathew Bullimore and Andrea Grigoletto for helpful discussions and useful comments on the draft.
\end{minipage}
\end{center}

\vspace{14pt}

\twocolumngrid

\let\oldaddcontentsline\addcontentsline
\renewcommand{\addcontentsline}[3]{}

 \bibliographystyle{JHEP}
 \small 
 \let\bbb\bibitem\def\bibitem{\itemsep5.5pt\bbb}
\bibliography{unitarity}

\let\addcontentsline\oldaddcontentsline

\end{document}